\documentclass[fleqn,usenatbib]{mnras}

\usepackage{newtxtext,newtxmath}

\usepackage[T1]{fontenc}

\DeclareRobustCommand{\VAN}[3]{#2}
\let\VANthebibliography\thebibliography
\def\thebibliography{\DeclareRobustCommand{\VAN}[3]{##3}\VANthebibliography}

\usepackage{graphicx}	
\usepackage{amsmath}	

\title[Early Excess Emission in SNe~Ia]{Possible Circumstellar Interaction Origin of the Early Excess Emission in Thermonuclear Supernovae} 

\author[M. Hu et al.]{
Maokai Hu,$^{1}$\thanks{kaihukaihu123@pmo.ac.cn}
Lifan Wang,$^{2}$\thanks{lifan@tamu.edu}
Xiaofeng Wang$^{3,4}$
and Lingzhi Wang$^{5,6}$
\\
$^{1}$Purple Mountain Observatory, Chinese Academy of Sciences, Nanjing 210023, China\\
$^{2}$Mitchell Institute for Fundamental Physics and Astronomy, Texas A$\&$M University, College Station, TX 77843, USA\\
$^{3}$Physics Department and Tsinghua Center for Astrophysics (THCA), Tsinghua University, Beijing, 100084, China\\
$^{4}$Beijing Planetarium, Beijing Academy of Science and Technology, Beijing, 100044, China\\
$^{5}$Chinese Academy of Sciences, South America Center for Astronomy, National Astronomical Observatories, CAS, Beijing 100101, China \\
$^{6}$CAS Key Laboratory of Optical Astronomy, National Astronomical Observatories, Chinese Academy of Sciences, Beijing 100101, China
}

\date{Accepted XXX. Received YYY; in original form ZZZ}

\pubyear{2015}

\begin{document}
\label{firstpage}
\pagerange{\pageref{firstpage}--\pageref{lastpage}}
\maketitle

\begin{abstract}
Type Ia supernovae (SNe~Ia) arise from the thermonuclear explosion in binary systems involving carbon-oxygen white dwarfs (WDs). The pathway of WDs acquiring mass may produce circumstellar material (CSM). Observing SNe~Ia within a few hours to a few days after the explosion can provide insight into the nature of CSM relating to the progenitor systems. In this paper, we propose a CSM model to investigate the effect of ejecta$-$CSM interaction on the early-time multi-band light curves of SNe~Ia. By varying the mass-loss history of the progenitor system, we apply the ejecta$-$CSM interaction model to fit the optical and ultraviolet (UV) photometric data of eight SNe~Ia with early excess. The photometric data of SNe~Ia in our sample can be well-matched by our CSM model except for the UV-band light curve of iPTF14atg, indicating its early excess may not be due to the ejecta$-$CSM interaction. Meanwhile, the CSM interaction can generate synchrotron radiation from relativistic electrons in the shocked gas, making radio observations a distinctive probe of CSM. The radio luminosity based on our models suggests that positive detection of the radio signal is only possible within a few days after the explosion at higher radio frequencies (e.g., $\sim250\ \text{GHz}$); at lower frequencies (e.g., $\sim1.5\ \text{GHz}$) the detection is difficult. These models lead us to conclude that a multi-messenger approach that involves UV, optical, and radio observations of SNe~Ia a few days past explosion is needed to address many of the outstanding questions concerning the progenitor systems of SNe~Ia. 
\end{abstract}




\section{Introduction}

Type Ia supernovae (SNe~Ia) are employed as the standardized candle in measuring cosmological distance through the luminosity-width relation \citep{1998AJ....116.1009R,2007ApJ...659...98R,1999ApJ...517..565P}, although their progenitor systems are still unclear (e.g., \citealt{2011NatCo...2..350H,2014ARA&A..52..107M}) and they may have different progenitor populations even for spectroscopically normal ones (e.g., \citealt{2013Sci...340..170W, 2019ApJ...882..120W}). The conventional scenario is that SNe~Ia are the results of the thermonuclear explosions of carbon-oxygen white dwarfs (WDs) whose masses approach the Chandrasekhar limit through merging with or accretion from a binary companion (e.g., \citealt{2000ARA&A..38..191H}). In the merger scenario, the so-called double degenerate (DD) channel, the companion is another carbon-oxygen WD \citep{1984ApJS...54..335I,1984ApJ...277..355W}, while in the single degenerate (SD) channel, a WD accretes matter from a main sequence, red giant, or Helium star \citep{1973ApJ...186.1007W,1982ApJ...253..798N}. These two channels may both encounter difficulties when confronted with observations. The DD channel predicts a relatively high degree of polarization \citep{2016MNRAS.455.1060B}, while the observed continuum polarization of SNe~Ia is usually lower than $0.2\%$ \citep{1997ApJ...476L..27W,doi:10.1146/annurev.astro.46.060407.145139,2016ApJ...828...24P,2019MNRAS.490..578C,2020ApJ...902...46Y}. On the other hand, direct evidences of the SD channel have not been found from extensive observational efforts, such as the null detection of H/He emission lines in the nebular spectrum \citep{2005A&A...443..649M,2013MNRAS.435..329L,2013ApJ...762L...5S,2016MNRAS.457.3254M,2020MNRAS.493.1044T}, and the absence of super-soft X-ray signals as can be expected from the accretion process of progenitors \citep{2008MNRAS.388..487N,2018MNRAS.481.4123K}. 

Multi-band observations within a few days after the explosion provide a powerful probe to investigate the physical origins of SNe~Ia. In the SD channel, interaction with the companion can lead to radiations in the X-ray, ultra-violet (UV), and optical wavelengths several hours after the explosion in certain viewing angles \citep{2010ApJ...708.1025K,2014ApJ...794...37M}. An early flux excess can also be produced if $^{56}$Ni is mixed to the outer layers of the ejecta due to hydrodynamic turbulence during the thermonuclear explosion \citep{2018A&A...614A.115M,2020A&A...634A..37M}, or if there is nuclear burning on the surface of the WD progenitor \citep{2017Natur.550...80J,2018ApJ...865..149J,2018ApJ...861...78M,2021ApJ...906...99L,2021MNRAS.502.3533M}. The interaction with circumstellar matter (CSM) can transform the kinetic energy of the ejecta into radiation and power the light curves of SNe with significant mass-loss history \citep{1994ApJ...420..268C,Wood-Vasey:2004ApJ...616..339W,2012ApJ...759..108S,2013MNRAS.435.1520M,2020PASJ...72...67T}. CSM interaction can also be the energy source of the first light curve peak shown a few days after the explosion for some core collapse supernovae \citep{2013ApJ...767..143B,2015ApJ...808L..51P,2018NatAs...2..808F,2021ApJ...910...68J}. Likely, the possibility exists that the early flux excess of SNe~Ia may originate from ejecta$-$CSM interaction \citep{2023MNRAS.522.6035M}.

In recent decades, a large amount of photometric and spectroscopic observations of SNe~Ia are available due to the rapid growth in time-domain surveys (e.g., \citealt{2001ASPC..246..121F,2009PASP..121.1395L,2017PASP..129j4502K,2019PASP..131g8001G}), but data within the first few days after the explosion are still rare. This situation is mainly limited by the cadence of the supernova survey program, which is usually around $2\sim3$ days to cover as large a survey area as possible. With recent wide-field supernova survey programs \citep{2009PASP..121.1395L,ATLAS2018PASP..130f4505T,ZTF2020PASP..132c8001D}, more and more early signals of SNe~Ia have been captured, such as the spectroscopic normal ones (e.g., SN~2011fe \citep{2011Natur.480..344N}, SN~2012cg \citep{2016ApJ...820...92M}, SN~2017cbv \citep{2017ApJ...845L..11H,2020ApJ...904...14W}, SN~2018oh \citep{2019ApJ...870L...1D,2019ApJ...870...12L}, SN~2019np \citep{2022MNRAS.514.3541S}, and SN~2021aefx \citep{2022ApJ...932L...2A,2022ApJ...933L..45H}), subluminous 2002es-like ones (e.g., iPTF14atg \citep{2015Natur.521..328C} and SN~2019yvq \citep{2020ApJ...898...56M,2021ApJ...919..142B}), and the super-Chandrasekhar explosion (e.g., SN~2020hvf \citep{2021ApJ...923L...8J}). 

The above nine SNe~Ia also constitute the sample of this paper. The first detection of SN~2011fe is just several hours after its explosion, and such early photometric data in consistence with a $t^{\alpha}$ law constrains the radius of the progenitor to that of a WD \citep{2011Natur.480..348L,2011Natur.480..344N,2012ApJ...744L..17B}. The other eight SNe~Ia are revisited in this paper because they show apparent flux excess during their early phases compared with the light curve of typical objects such as SN~2011fe. In particular, SN~2017cbv exhibits apparent blue excess in its early phases. This flux excess may be generated from the decay of $^{56}$Ni mixed in the outer layers of the ejecta \citep{2020A&A...642A.189M}, $^{56}$Ni produced at the surface layers due to a Helium detonation \citep{2018ApJ...861...78M}, the interaction with the companion star \citep{2017ApJ...845L..11H}, or ejecta$-$CSM interaction. For the companion interaction scenario, the predicted large amount of UV radiation is not supported by observations \citep{2017ApJ...845L..11H}. Besides, the predicted H/He emission lines in the nebular spectra relating to the SD channel are not observed for SN~2017cbv \citep{2018ApJ...863...24S}. 

In this paper, we revisited the influence of CSM interaction on the early multi-band light curves of SNe~Ia, since the popular channels of progenitor systems may generate CSM through the processes involving mass accretion/excretion, stellar wind, or nova explosions. Section~\ref{section22} describes the early flux excess of the eight revisited SNe~Ia in our sample. In Section~\ref{section33}, two models of ejecta$-$CSM interaction are introduced. The fits to the optical and UV luminosity are shown in Section~\ref{section44}. We show the radio radiation from the relativistic electrons generated by the ejecta$-$CSM interaction in Section~\ref{section55}. The conclusions are given in Section~\ref{section66}.

\section{The Early Excess of Thermonuclear Supernovae} 
\label{section22}  

\begin{figure}
\centering
\includegraphics[width = 0.95 \linewidth]{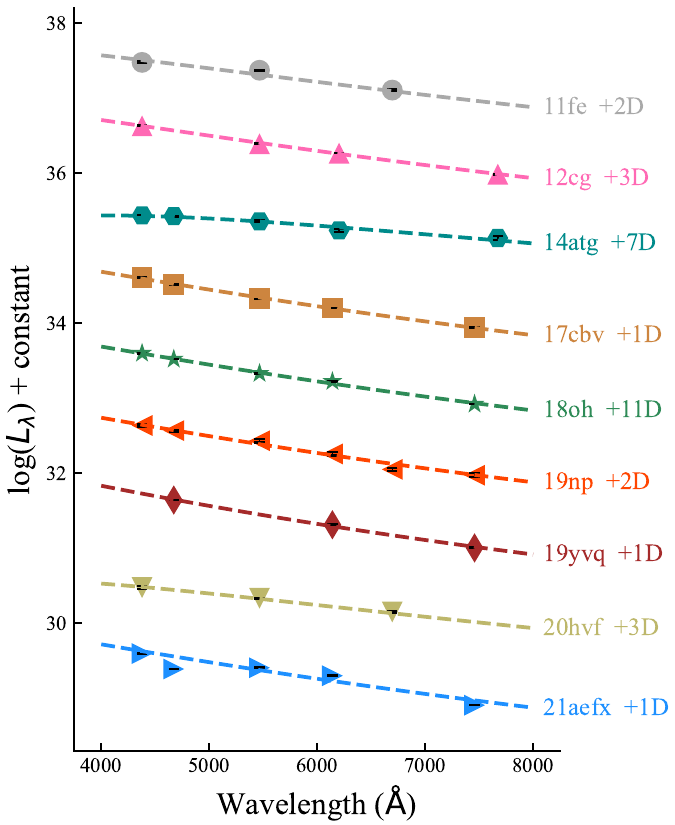}
\caption{The symbols are the early-phase luminosity of the revisited SNe~Ia at each optical band's efficient wavelength ($L_{\lambda}$). The dashed lines are the corresponding black-body spectrum fits. The phases listed in the figure correspond to the explosion of each SN~Ia.} 
\label{fig_00} 
\end{figure} 

Several recent studies have modeled the early-phase observations of SNe~Ia through their UV properties \citep{2012ApJ...749...18B}, optical rises \citep{2020ApJ...892...25J,2020ApJ...902...47M,2021ApJ...908...51F}, and color evolutions \citep{2020ApJ...902...48B}. In this paper, we focus on the ejecta$-$CSM interaction to model eight SNe~Ia with the strongest evidences of early flux excess. Among them, SN~2012cg \citep{2016ApJ...820...92M}, iPTF14atg \citep{2015Natur.521..328C}, and SN~2019yvq \citep{2020ApJ...898...56M} show an initial declining flux excess in the UV bands which may be related to the ejecta$-$CSM interaction. The early flux excesses of SN~2017cbv \citep{2017ApJ...845L..11H,2018ApJ...861...78M,2020A&A...642A.189M}, SN~2018oh \citep{2019ApJ...870L...1D,2019ApJ...872L...7L}, SN~2019np \citep{2022MNRAS.514.3541S}, and SN~2021aefx \citep{2022ApJ...932L...2A,2022ApJ...933L..45H} are still under debate, while SN~2020hvf \citep{2021ApJ...923L...8J} seems to show optical bumps within the first day since the discovery which is consistent with the expectations from ejecta$-$CSM interaction. SN~2016jhr also has early observations showing flux excess compared to typical normal SNe~Ia, but it is not in our sample since its early flash is likely to be triggered by a helium detonation on the surface of the WD \citep{2017Natur.550...80J}. 

\begin{figure*}
\centering
\includegraphics[width = 0.9 \linewidth]{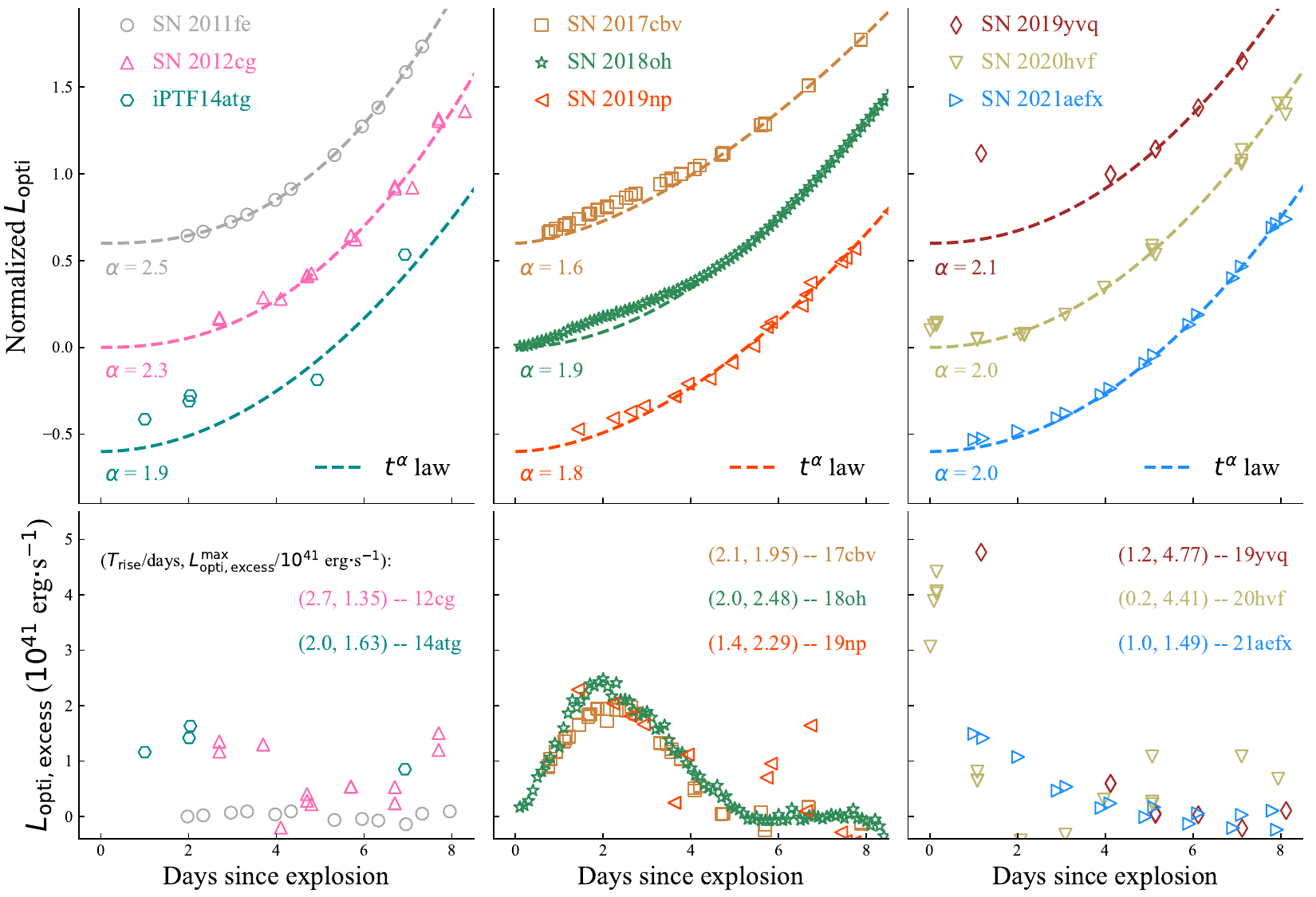}
\caption{Upper panels: the symbols represent the normalized optical luminosity ($L_{\text{opti}}$) of SN~2011fe \citep{2016ApJ...820...67Z} which has no early excess and the eight SNe~Ia with early excess, including SN~2012cg \citep{2016ApJ...820...92M}, iPTF14atg \citep{2015Natur.521..328C}, SN~2017cbv \citep{2017ApJ...845L..11H,2020ApJ...904...14W}, SN~2018oh \citep{2019ApJ...870L...1D,2019ApJ...870...12L}, SN~2019np \citep{2022MNRAS.514.3541S}, SN~2019yvq \citep{2020ApJ...898...56M,2021ApJ...919..142B}, SN~2020hvf \citep{2021ApJ...923L...8J} and SN~2021aefx \citep{2022ApJ...932L...2A,2022ApJ...933L..45H}. $L_{\text{opti}}$ is the integration of the black-body spectrum from 4000 $\text{\AA}$ to 8000 $\text{\AA}$ fitted by their multi-band photometric data except for iPTF14atg, SN~2018oh, and SN~2020hvf due to the lack of multi-band observations during their early phases (see Section~\ref{section22} for more details). The dashed lines show the $t^{\alpha}$ fit to the photometric data of each SN~Ia. The lower panels show the differences between the $L_{\text{opti}}$ of each SN~Ia and the $t^{\alpha}$ law derived from the SN~Ia itself. For comparison, we list the rising time ($T_{\text{rise}}$) and the peak of the $L_{\text{opti, excess}}$ ($L^{\text{max}}_{\text{opti, excess}}$) of the eight SNe~Ia with early excess.  } 
\label{fig_11} 
\end{figure*} 

The optical light curves of SNe~Ia in our studies have different photometric systems, including the Sloan Digital Sky Survey photometry, the Johnson-Cousins $UBVRI$ system, the Kepler filter (SN~2018oh), and no-filter observations (SN~2020hvf). All the UV-band light curves are from the Swift satellite. Therefore, we adopt the optical luminosity ($L_{\text{opti}}$) to characterize the early excess of the eight SNe~Ia in our study to reduce the influence of the magnitude systems among the observations of SNe~Ia, and we adopt the UV-band luminosity ($L_{\text{UV}}$) to represent the early-phase evolution in UV bands. The $L_{\text{opti}}$ defined in this paper is the integration of the black-body spectrum fitted by the multi-band photometric data from 4000 $\text{\AA}$ to 8000 $\text{\AA}$. As shown in Figure~\ref{fig_00}, the black-body spectrum is a favorable profile fitting the early-time multi-band photometric data of SNe~Ia. For iPTF14atg, SN~2018oh, and SN~2020hvf, the optical multi-band observations are absent in the duration of early excess, and the observational band is PTF$r$ band (iPTF14atg), Kepler filter (SN~2018oh), or no-filter (SN~2020hvf), respectively. Therefore, we adopted the shape of the single-band light curve as the shape of the $L_{\rm opti}$ curve during the flux-excess phases for these three SNe~Ia. We then shifted the single-band flux to the scale of the corresponding optical luminosity with the overlap between the single-band data and multi-band data to obtain the $L_{\rm opti}$ curve.

The optical photometric data of SNe~Ia are de-reddened from the extinction of the Milky Way and host galaxies. The color excess ($E(B-V)$), the total-to-selective extinction ratio ($R_V$), and the luminosity distance of the SNe~Ia are all from their respective references. Note that we adopted 12.3 Mpc as the distance of SN~2017cbv derived from \citet{2018ApJ...863...24S}. A similar distance to SN~2017cbv is also adopted in several other studies \citep{2018ApJ...863...90W,2020ApJ...895..118B,2020ApJ...904...14W}. Figure~\ref{fig_11} displays the normalized $L_{\text{opti}}$ curves of the eight SNe~Ia, together with SN~2011fe for comparison.
We adopt a modified fireball model to fit the early-time luminosity of SNe~Ia, in which $L_{\rm opti} \propto t^{\alpha}$, where $t$ is the time since the explosion, and $\alpha$ is a power-law index \citep{1999AJ....118.2675R,2011MNRAS.416.2607G}. Note that the 'explosion time' defined in our paper is likely the first-light time due to the possible existence of a so-called 'dark phase' between the explosion epoch and the time of the first time for SNe~Ia. The index $\alpha$ is not fixed to be $2.0$ owning to the possible evolution of expansion velocity or fireball temperature during the early time. Besides, we fitted the early-time luminosity curve using the epochs from +5 days to +8 days since the explosion due to the early excess of the revisited SNe~Ia in our study. The ratio between the optical luminosity at +8 days since the explosion and the peak luminosity is about 0.4, consistent with the choice in \citet{2020ApJ...902...47M}.
The early-time $L_{\text{opti}}$ curve of SNe~Ia satisfies the $t^{\alpha}$ law with the index $\alpha \sim 2.0$, which is consistent with previous results \citep{2013A&A...554A..27P,2015MNRAS.446.3895F,2016ApJ...820...67Z}. The early-time optical excesses of the eight SNe~Ia over the $t^{\alpha}$ law ($L_{\text{opti, excess}}$) are shown in the lower panel of Figure~\ref{fig_11}, and it can be roughly described by two quantities, the maximum of $L_{\text{opti, excess}}$ ($L_{\text{opti, excess}}^{\text{max}}$) and the rising time ($T_{\text{rise}}$) of $L_{\text{opti, excess}}$ since the explosion. For simplicity, the values of $L_{\text{opti, excess}}^{\text{max}}$ and $T_{\text{rise}}$ are just from the corresponding data point without any process like the Gaussian process fit or smooth process. These two quantities can provide a preliminary diagnosis of our ejecta$-$CSM interaction model.

\begin{figure*}
\centering
\includegraphics[width = 0.9 \linewidth]{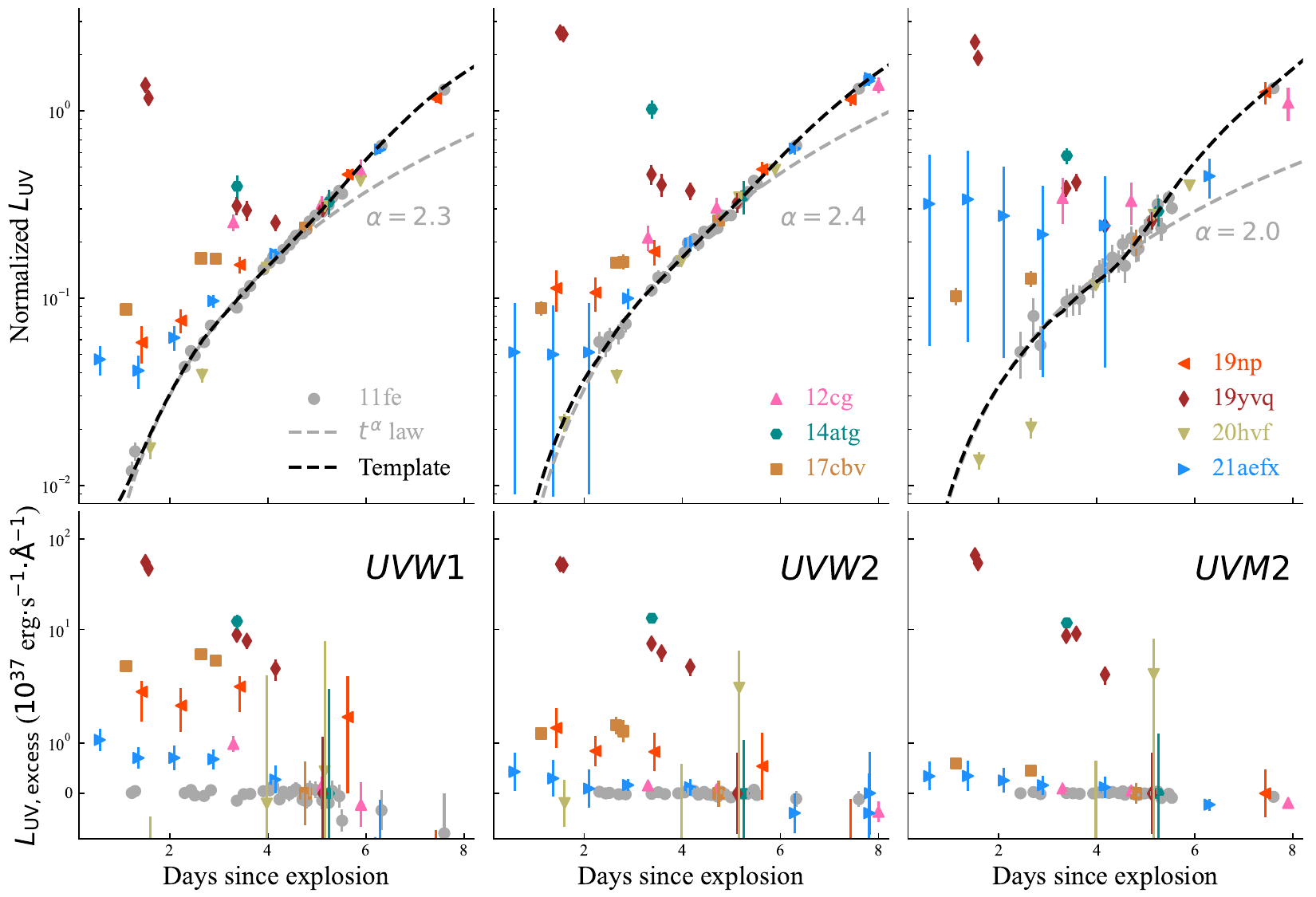}
\caption{Be similar with Figure~\ref{fig_11} but for the luminosity of $UVW1$ band, $UVW2$ band, and $UVM2$ band, respectively. The dashed gray lines are the $t^{\alpha}$ fit to the photometric data of SN~2011fe. The dashed black lines are smooth to the photometric data of SN~2011fe and are regarded as the template of UV-band luminosity in our study. The lower panel shows the differences between the $L_{\text{UV}}$ of each SN~Ia and the template of $L_{\text{UV}}$ derived from SN~2011fe. The references of the photometric data are same as shown in Figure~\ref{fig_11} except for SN~2011fe \citep{2012ApJ...753...22B}. } \label{fig_22} 
\end{figure*}

The same process is also applied to generate the early-phase UV-band light curves of each SN~Ia with extinction corrections using the same $R_V$ and $E(B-V)$ as for $L_{\text{opti}}$. Figure~\ref{fig_22} shows the normalized $L_{\text{UV}}$ of SNe~Ia. Similarly, a $t^{\alpha}$ law of $L_{\text{UV}}$ is generated from the observed data of SN~2011fe, with $\alpha = 2.3$, 2.4, or 2.0 for $UVW1$, $UVW2$, or $UVM2$ band, which are consistent with the result reported in \citet{2016ApJ...820...92M}. Note that the template of the UV-band light curve is from the smoothed curve of SN~2011fe rather than the fitted $t^{\alpha}$ law of SN~2011fe, and the difference between the smoothed curve and the $t^{\alpha}$ law is minimal within a few days since the explosion as shown in Figure~\ref{fig_22}. Comparing Figure~\ref{fig_11} and Figure~\ref{fig_22}, SN~2012cg, iPTF14atg, SN~2017cbv, SN~2019np, SN~2019yvq, and SN~2021aefx all have early-time multi-band observations (from optical to UV bands), and all show significant excess over the $t^{\alpha}$ law, while the UV-band coverage is absent for SN~2018oh and SN~2020hvf during the phases corresponding to the early optical excess.

With the definition of both $L_{\text{opti}}$ and $L_{UVW1}$, it is straightforward to examine the possible CSM interaction origin of the early excess emission in SNe~Ia. We will compare $L_{\text{opti, excess}}^{\text{max}}$ and $T_{\text{rise}}$ of the revisited SNe~Ia with our CSM model covering a broad range of model parameters to give a quick look of whether our CSM model is reasonable for the early excess of SNe~Ia. We then fit the early excess of $L_{\text{opti}}$ curves and predict the related $L_{UVW1}$ curves. The model parameters from these well-fitted models are employed to predict further the radio radiations related to the ejecta$-$CSM interactions of these SNe~Ia. 

\section{The CSM Interaction Model} 
\label{section33} 

The ejecta$-$CSM interaction has been studied previously for SNe (e.g., \citealt{1982ApJ...258..790C,1994ApJ...420..268C,Wood-Vasey:2004ApJ...616..339W,2013MNRAS.435.1520M}). The CSM density ($\rho_{\text{csm}}$) considered in this study follows the expression  $\rho_{\text{csm}} = \dot{M}_{\text{w}}/(4\pi R^2 v_{\rm w})$, where $R$ is the distance from the SN, $\dot{M}_{\text{w}}$ is the mass-loss rate of CSM, and $v_{\rm w}$ is the wind speed. We adopt $v_{\text{w}} = 10\ \text{km}\cdot{\rm s}^{-1}$ in the study. For a constant $\dot{M}_{\text{w}}$, the total CSM mass $M_{\text{csm}}$ is equal to $(R_{\text{out}} - R_{\text{in}})\dot{M}_{\text{w}} / v_{\rm w}$ with $R_{\text{in}}$ and $R_{\text{out}}$ being the inner and outer boundaries of the CSM, respectively. In general, $R_{\text{in}}$ is related to the position of the surface of the progenitor and we set $R_{\rm in}$ to zero, while $R_{\text{out}}$ is assumed to vary from $10^{11}\ \text{cm}$ to $10^{16}\ \text{cm}$ in our study.

The velocity of SN ejecta ($v_{\text{ej}}$) satisfies $v_{\text{ej}} = R/t$ as expected from a homologous expansion, where $t$ is the time since SN explosion. The density ($\rho_{\text{ej}}$) of the ejecta follows the power-law profile of $\rho_{\text{ej}} \propto R^{-\delta}$ and $\rho_{\text{ej}} \propto R^{-n}$ for regions interior and exterior of a transition velocity $v_t$ \citep{1999ApJ...510..379M,2010ApJ...708.1025K}, respectively. The indices $n$ and $\delta$ are equal to $10.0$ and $0.5$ as expected from self-similar solutions. The transition velocity $v_t$ is formulated by the SN kinetic energy $E_{\text{ej}}$ and ejecta mass $M_{\text{ej}}$ as $v_t = [\frac{2(5-\delta)(n-5)E_{\text{ej}}}{(3-\delta)(n-3)M_{\text{ej}}}]^{1/2}$ \citep{2013MNRAS.435.1520M}. The value of $v_t$ is about $1.2\times10^4\ \text{km}\cdot{\rm s}^{-1}$ assuming $E_{\text{ej}} = 1.5\times10^{51}\ \text{erg}$ and $M_{\text{ej}} = 1.4\ \text{M}_{\odot}$ \citep{2018ApJ...861...78M}.

\begin{figure}
\centering
\includegraphics[width = 0.9 \linewidth]{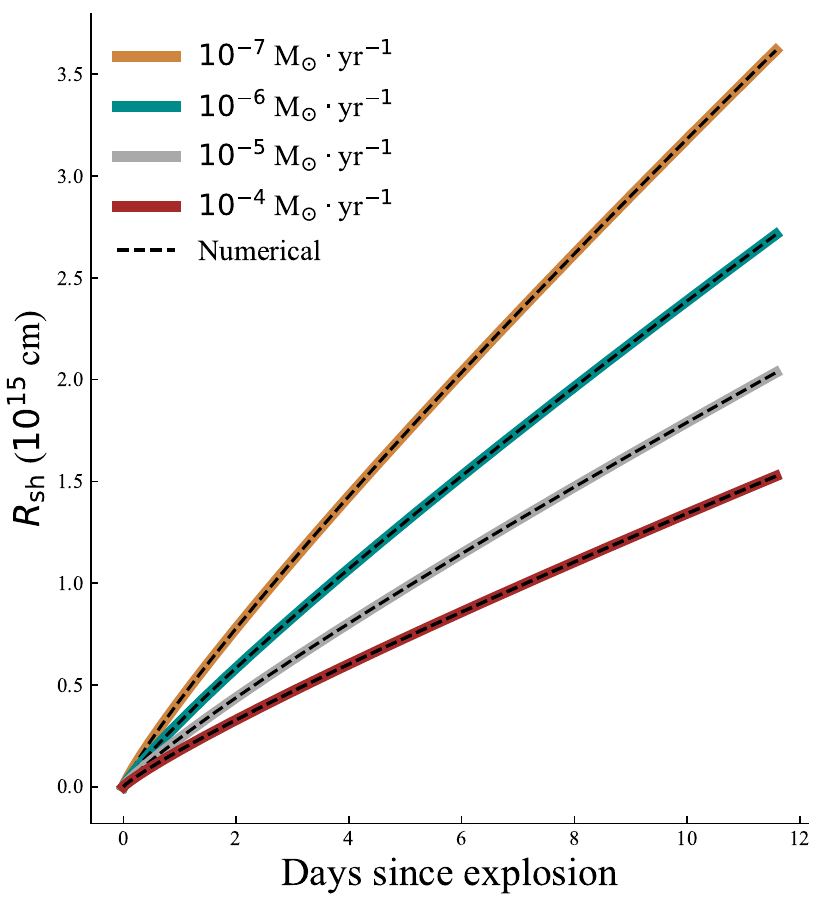}
\caption{The orange, cyan, gray, and red solid lines show the evolution of the radii of the shocked CSM shell ($R_{\text{sh}}$)  calculated by the formula in \citet{2013MNRAS.435.1520M} with $\dot{M}_{\text{w}}$ of $10^{-7}\ \text{M}_{\odot}\cdot\text{yr}^{-1}$, $10^{-6}\ \text{M}_{\odot}\cdot\text{yr}^{-1}$, $10^{-5}\ \text{M}_{\odot}\cdot\text{yr}^{-1}$, and $10^{-4}\ \text{M}_{\odot}\cdot\text{yr}^{-1}$, respectively. The dashed black lines are the evolution of $R_{\text{sh}}$ solved numerically by our CSM model.} 
\label{fig_33} 
\end{figure}

\subsection{Model\_sh} 

The first scenario considered in our study, named Model\_sh, has the characteristic parameters of $R_{\text{out}}\sim10^{12}\ \text{cm}$, $\dot{M}_{\text{w}}\sim10^{-1}\ \text{M}_{\odot}\cdot\text{yr}^{-1}$, and the corresponding total CSM mass $M_{\text{csm}}\sim0.003\ \text{M}_{\odot}$. The duration of CSM interaction in the Model\_sh is less than an hour, and the interaction process can be regarded as a shock breakout, which results in a thin shell expanding with velocity $V_{\text{sh}}$ at a distance $R_{\text{sh}}$ and a shell thickness $\Delta R_{\text{sh}}$. We adopt $\Delta R_{\text{sh}}/R_{\text{sh}}\sim0.2$ in the Model\_sh, which is different from \citet{2018ApJ...861...78M}, but is consistent with the results in \citet{1982ApJ...258..790C}. The $R_{\text{sh}}$ evolves as $R_{\text{sh}} = R_{\text{out}} + V_{\text{sh}}t$. $V_{\text{sh}}$ is determined by the equation assuming that the mass of the shocked ejecta is equal to the total mass of the CSM such that $\int_{V_{\text{sh}}}^{\infty}4\pi (vt)^2\rho_{\text{ej}}t\mathrm{d}v = M_{\text{csm}}$ \citep{2018ApJ...861...78M}. The bolometric luminosity ($L$) from this adiabatically expanding shell can be solved by the first law of thermodynamics as $L\propto\exp(-\frac{t_ht+t^2/2}{t_ht_d(0)})$, where $t_h = R_{\text{out}} / V_{\text{sh}}$ and $t_d(0)$ is the diffusion timescale when $t = 0$ \citep{2018ApJ...861...78M}. The observed multi-band light curves can be generated with the assumption of blackbody radiation. Note that the bolometric luminosity is monotonically decreasing with time, while the light curve of a certain waveband has a unimodal structure. Thus the predicted flux contributions by Model\_sh allow calculations of quantities such as the maximum optical luminosity of the ejecta$-$CSM interaction and the rising time since the explosion.

\subsection{Model\_ext} 

The interaction with extended CSM cannot be simplified to the shock breakout process since the interaction can last more than a few days. A similar situation may happen for SNe~Ia, because the mass-loss history for the progenitor may be long enough to generate CSM with an extended distribution. Based on this picture, we consider the scenario Model\_ext, which has a more extended CSM (e.g., the outer boundary of CSM is $\sim10^{15}$ cm), and we assume that the un-shocked CSM is optically thin. The evolution of $R_{\text{sh}}$ and $V_{\text{sh}}$ for the shocked CSM satisfies the conservation of momentum as follows, 
\begin{equation} 
\label{eq1} 
M_{\text{sh}}\frac{\mathrm{d}V_{\text{sh}}}{\mathrm{d}t} = 4\pi R_{\text{sh}}^2[\rho_{\text{ej}}(v_{\text{ej}} - V_{\text{sh}})^2 - \rho_{\text{csm}}(V_{\text{sh}} - v_{w})^2]
\end{equation} 
where $M_{\text{sh}}$ is the total mass of the shocked ejecta and CSM. In  Model\_ext, we only consider the interaction process during the first few days after the explosion, and $\dot{M}_{\text{w}}$ is basically less than $10^{-4}\ \text{M}_{\odot}\cdot\text{yr}^{-1}$ as has been constrained by  radio or X-ray observations of SNe~Ia \citep{2006ApJ...646..369P,2012ApJ...748L..29R,2016ApJ...821..119C,2020ApJ...890..159L}. Thus, the shocked SN ejecta is always confined inside the exterior part of the ejecta with $v_{\text{ej}} > v_t$. With the solution of the kinetic evolution, the corresponding bolometric luminosity $L$ is given by the power of the shocked CSM with a conversion efficiency $\epsilon$ as $L = \frac{\epsilon}{2}\dot{M}_{\text{w}}V_{\text{sh}}^3$, where $\epsilon = 0.15$ in our simulations in consistence with previous studies \citep{1982ApJ...258..790C,2013MNRAS.435.1520M}. On the other hand, one important quantity in the Model\_ext is $\dot{M}_{\text{w}}(R)$ which is a function of the distance $R$ as given below, 
\begin{equation} 
\label{eq2}
\dot{M}_{\text{w}}(R) = \begin{cases} 
\dot{M}_{\text{w}}(0)(\frac{R}{R_1})^{n_1}, & R \le R_1 \\ 
\dot{M}_{\text{w}}(0), & R_1 < R \le R_2 \\
\dot{M}_{\text{w}}(0) (\frac{R_3 - R}{R_3 - R_2})^{n_2}, & R_2 < R \le R_3 \\
\end{cases}
\end{equation} 
As shown in Equation~\ref{eq2}, $\dot{M}_{\text{w}}(R)$ increases to $\dot{M}_{\text{w}}(0)$ within the distance of $R_1$ relating to an index of $n_1$. $\dot{M}_{\text{w}}(R)$ equals to a constant $\dot{M}_{\text{w}}(0)$ between $R_1$ and $R_2$. $\dot{M}_{\text{w}}(R)$ decreases to zero from $R_2$ to $R_3$ with an index of $n_2$. CSM could be ignored for a distance larger than $R_3$. The range of parameters $n_1$ and $n_2$ is from 0.0 to 3.0.

Therefore, the observed light curves for Model\_ext can be numerically solved based on Equation~\ref{eq1}. As a simplified situation with a constant $\dot{M}_{\text{w}}(R)$, \citet{2013MNRAS.435.1520M} acquired the integrated formula of the luminosity curve of CSM interaction. We compared the evolution of $R_{\text{sh}}$ between the integrated formula from \citet{2013MNRAS.435.1520M} and our numerical solutions with a constant $\dot{M}_{\text{w}}(R)$ as shown in Figure~\ref{fig_33}, which demonstrate the validity of our numerical procedure.
Note that our ejecta$-$CSM interaction model is one-dimension assuming a spherical explosion and spherical distribution of CSM. The polarimetric observations suggest the explosion of SNe~Ia is approximately spherical \citep{doi:10.1146/annurev.astro.46.060407.145139}, and the distribution of CSM is also spherical if the mass-loss process is isotropic. However, the interaction with an aspherical CSM may exist, which may generate different luminosity curves and possible polarimetric signals.

\begin{figure} 
\centering 
\includegraphics[width = 0.9\linewidth]{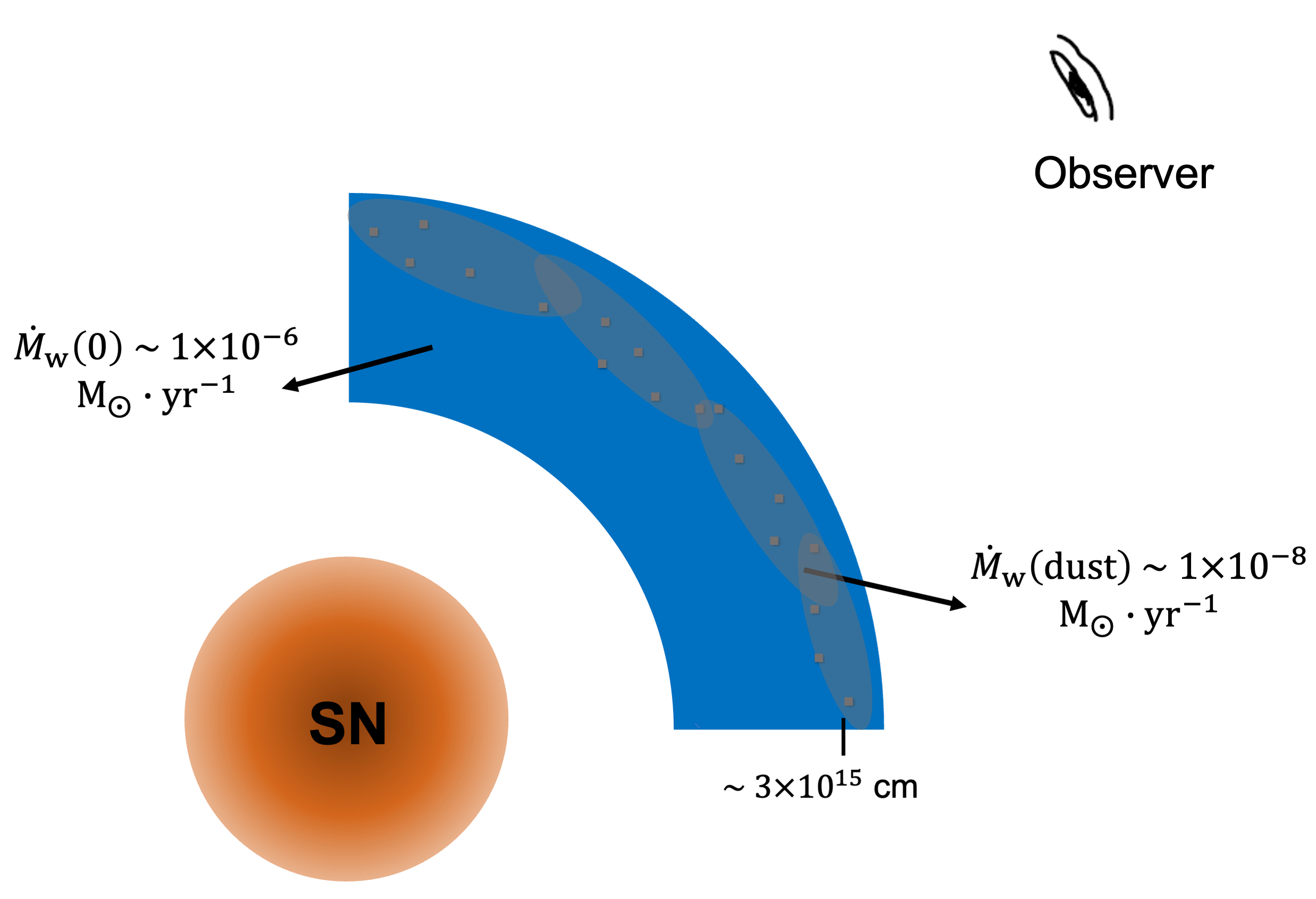}
\caption{The illustration of the configuration of Model\_ext with dusty CSM around SNe~Ia modeling UV-band light curves. The gray area with gray dots is where the dust exists. The mass-loss rate of the dust ($\dot{M}_{\text{w}}(\text{dust})$) is related to $B-$band optical depth of 0.15 and the dust distance of $3\times10^{15}$ cm. For comparison, $\dot{M}_{\text{w}}(0)$ with a characteristic value of around $1\times10^{-6}\ \text{M}_{\odot}\cdot\text{yr}^{-1}$ is also shown in the figure.}
\label{fig_an_ilus} 
\end{figure}

\begin{figure} 
\centering
\includegraphics[width = 0.95 \linewidth]{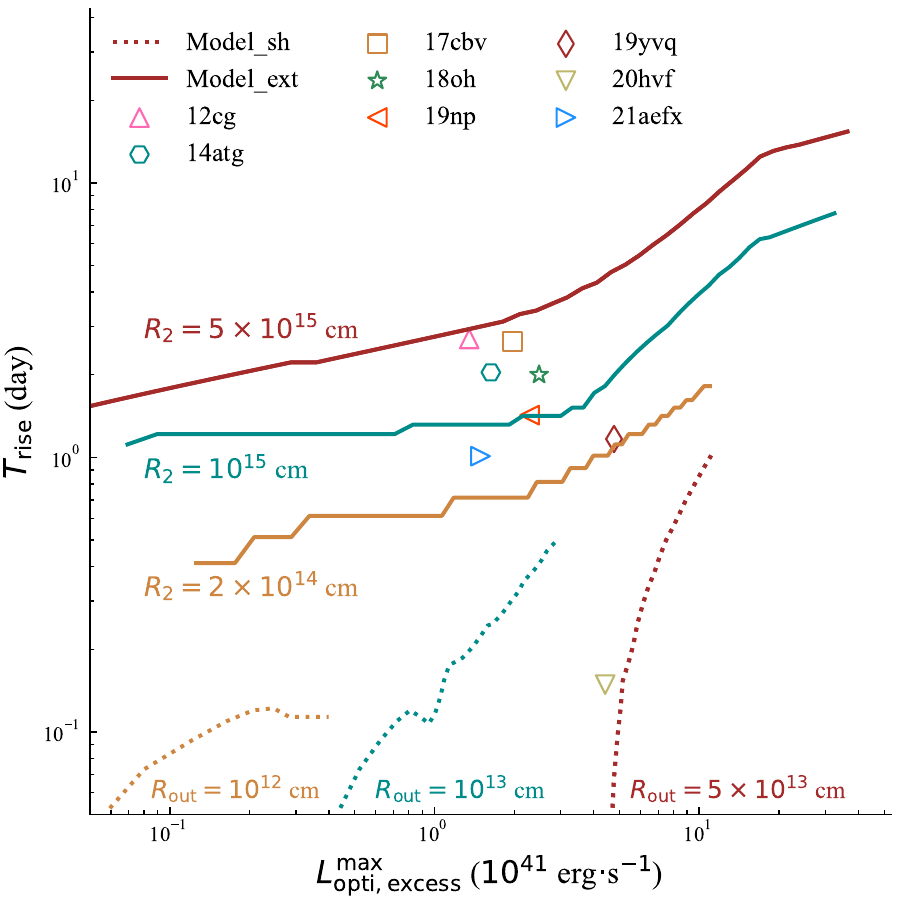}
\caption{The lines are the predicted rising time of the optical excess versus the maximum of the optical excess for Model\_sh (dotted lines) with $R_{\text{out}}$ of $10^{12}\ \text{cm}$ (yellow), $10^{13}\ \text{cm}$ (cyan), and $5\times10^{13}\ \text{cm}$ (red), and for Model\_ext (solid lines) with $R_{2}$ of $2\times10^{14}\ \text{cm}$ (yellow), $10^{15}\ \text{cm}$ (cyan), and $5\times10^{15}\ \text{cm}$ (red), respectively. For each line of Model\_sh, the $\dot{M}_{\text{w}}$ ranges from $0.001$ to $1.0$ $\text{M}_{\odot}\cdot\text{yr}^{-1}$, and the value of $\dot{M}_{\text{w}}(0)$ for each line of Model\_ext ranges from $10^{-7}$ to $10^{-4}$ $\text{M}_{\odot}\cdot\text{yr}^{-1}$. For simplicity, we set $R_1 = 0.4\times R_2$ and $R_3 = 2.0\times R_2$.} All the symbols are the $T_{\text{rise}}$ and $L_{\text{opti, excess}}^{\text{max}}$ calculated from the luminosity residual shown in the lower panel of Figure~\ref{fig_11} for the eight revisited SNe~Ia in this paper. 
\label{fig_44} 
\end{figure}

\subsection{Dusty CSM}

Assuming any typical gas-to-dust ratios, we introduce the effect of dusty CSM on the UV-band light curves. As estimated in \citet{2011ApJ...735...20A}, the pre-existing circumstellar dust within $\sim 10^{16}\ {\rm cm}$ would be destroyed by the peak luminosity of SNe~Ia. Consequently, the evaporation radius of circumstellar dust would rapidly increase as the increase of bolometric luminosity soon after the explosion. Taking SN~2011fe as an example, the bolometric luminosity at +1, +2, and +4 days since the explosion is about $3.5\times10^{40}\ {\rm erg}\cdot{\rm s}^{-1}$, $2.0\times10^{41}\ {\rm erg}\cdot{\rm s}^{-1}$, and $1.4\times10^{42}\ {\rm erg}\cdot{\rm s}^{-1}$, respectively. With the assumption of the peak bolometric luminosity of $10^{43}\ {\rm erg}\cdot{\rm s}^{-1}$ and a rough estimation of the evaporation radius from \citet{2011ApJ...735...20A}, the hypothesized evaporation radii for SN~2011fe at +1, +2, and +4 days since the explosion is about $6\times10^{14}$ cm, $1.4\times10^{15}$ cm, and $3.7\times10^{15}$ cm, respectively. However, time-dependent dust destruction is a complicated process during the early phase of SNe~Ia.
Nevertheless, we only consider the dusty CSM in the Model\_ext rather than in Model\_sh due to the difference in characteristic distances. To investigate the absorption and scattering from circumstellar dust, we consider a simple dust model in which the chemical composition is just silicate with a typical size of $0.05\ \mu\text{m}$, indicating that the dust extinction is more significant in UV bands than in the optical. 

For simplicity, We assume a spherical distribution of the dust within an inner boundary of $1\times10^{15}\ \text{cm}$ and an outer boundary of $5\times10^{15}\ \text{cm}$. The optical depth in $B$-band is adopted as 0.15, the corresponding optical depth in $UVW1$ band is 1.1, and the averaged optical depth from 4000 $\text{\AA}$ to 8000 $\text{\AA}$ is about 0.07. Thus, the radiative transfer process in dusty CSM for optical bands is ignored in this paper. Assuming the same wind velocity ($10\ \text{km}\cdot{\rm s}^{-1}$), the mass-loss rate of the dust is about $1\times10^{-8}\ \text{M}_{\odot}\cdot\text{yr}^{-1}$, which is about $10^{-2}$ times of the typical value of $\dot{M}_{\text{w}}(0)$ ($\sim 10^{-6}\ \text{M}_{\odot}\cdot\text{yr}^{-1}$) in the Model\_ext as illustrated in Figure~\ref{fig_an_ilus}. 
We assume that the inner boundary of circumstellar dust increases linearly from $1\times10^{15}\ \text{cm}$ at the initial state to $5\times10^{15}\ {\rm cm}$ at +4 days since the explosion, which is consistent with the above discussion of the hypothesized evaporation radius relating to the early-phase bolometric luminosity of SN~2011fe.
The time-dependent dust destruction makes the radiative transfer in the dusty CSM a dynamic process. We incorporated this dynamic process in our Monte Carlo radiative transfer program in \citet{2022ApJ...931..110H} to solve for the UV fluxes in dusty CSM. 

\section{Fitting the Early Excess Emission with CSM Interaction}
\label{section44}  

Figure~\ref{fig_44} displays the predicted rising time of the optical excess versus the maximum of the optical excess for Model\_sh and Model\_ext with different parameter configurations. For Model\_sh, $R_{\text{out}}$ is set to $10^{12}\ \text{cm}$, $10^{13}\ \text{cm}$, and $5\times10^{13}\ \text{cm}$, and $\dot{M}_{\text{w}}$ is set to from $0.001\ \text{M}_{\odot}\cdot\text{yr}^{-1}$ to $1.0\ \text{M}_{\odot}\cdot\text{yr}^{-1}$. The corresponding total CSM mass is in the range of $1.6\times10^{-5}\ \text{M}_{\odot}$ to $0.16\ \text{M}_{\odot}$. For Model\_ext, although $R_1$, $R_2$, $R_3$, and $\dot{M}_{\text{w}}(0)$ are all free parameters, $R_2$ and $\dot{M}_{\text{w}}(0)$ can significantly influence the flux relating to the ejecta$-$CSM interaction. The ranges of parameter $R_2$ considered here is set to $2\times10^{14}\ \text{cm}$, $10^{15}\ \text{cm}$, and $5\times10^{15}\ \text{cm}$, and $\dot{M}_{\text{w}}(0)$ varies from $10^{-7}\ \text{M}_{\odot}\cdot\text{yr}^{-1}$ to $10^{-4}\ \text{M}_{\odot}\cdot\text{yr}^{-1}$. It is clearly shown that Model\_sh in each parameter grid has the characteristics of a very short duration, which is contradictory to the early flux excess of the revisited SNe~Ia in this paper except for SN~2020hvf. Meanwhile, Model\_ext with certain parameters can fit the early optical excess of SNe~Ia satisfactorily. However, combining the photometric data of optical and UV bands may examine the hypothesis that the early excess arises from the ejecta$-$CSM interaction.

\begin{figure*}
\centering
\includegraphics[width = 0.95 \textwidth]{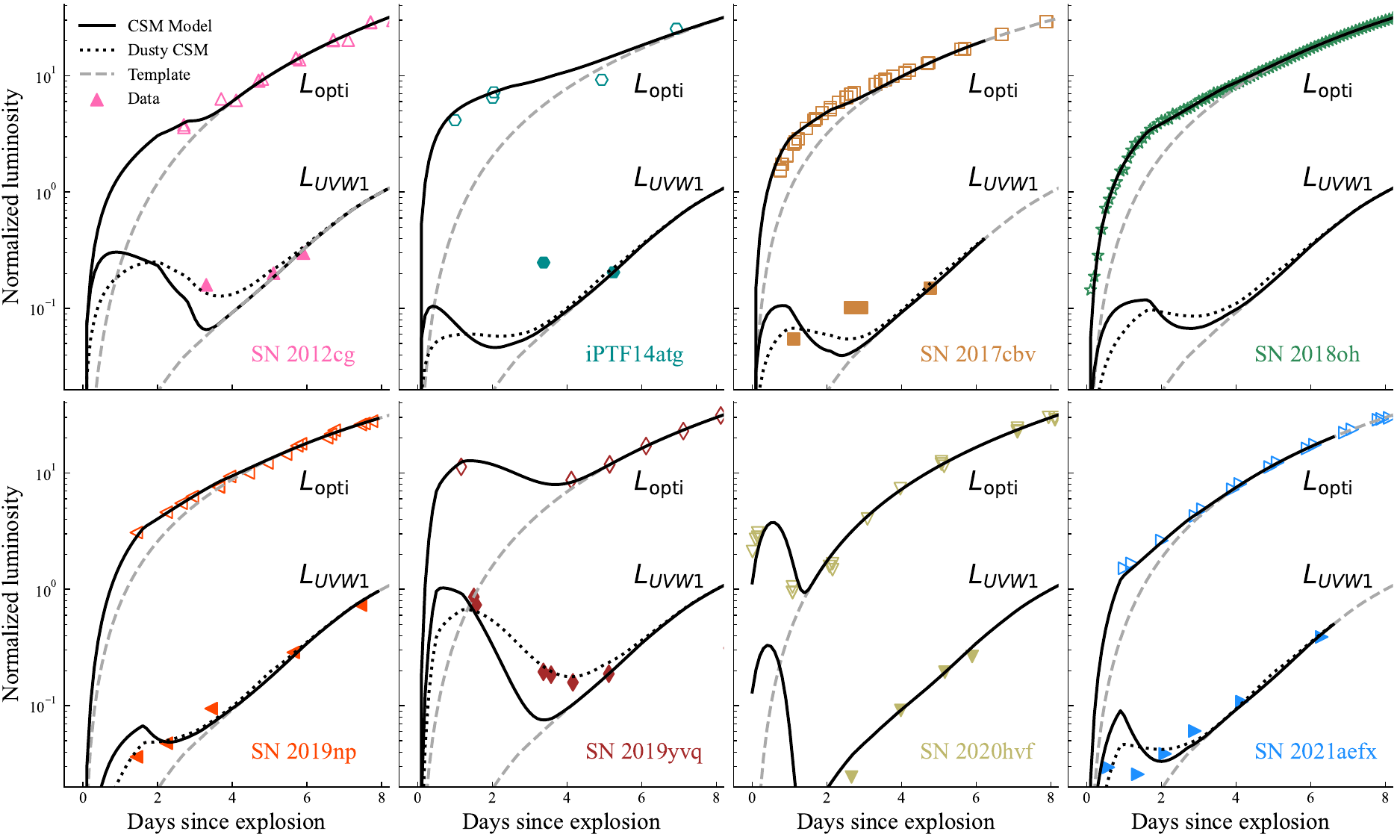}
\caption{The results of our CSM model fitting the early excess of the eight revisited SNe~Ia in this paper. Model\_sh is used to fit the signal of SN~2020hvf, and Model\_ext is used for the rest seven SNe~Ia. For each panel, the dashed gray lines are the normalized luminosity template generated from $t^{\alpha}$ law ($L{\text{opti}}$) or smoothing process ($L_{UVW1}$). The solid black lines are the fitted luminosity curves from our CSM model without considering the existence of dust, while the dotted black lines are the predicted $L_{UVW1}$ curves with the extinction and scattering of circumstellar dust. All the symbols are the data of each SN~Ia with the same color and shape as shown in Figure~\ref{fig_11}.} 
\label{fig_55} 
\end{figure*}

We adopt Model\_sh to fit the early-time optical excess of SN~2020hvf ($R_{\text{out}}=3\times10^{13}$ cm, $M_{\text{CSM}} = 0.05\ \text{M}_{\odot}$) and Model\_ext for the rest seven SNe~Ia with the parameter values shown in Table~\ref{para_model}. The fitted optical luminosity curves and the predicted $UVW1$-band luminosity are shown in Figure~\ref{fig_55}. The result clearly suggests that ejecta$-$CSM interaction can explain the early excess in the optical band of SNe~Ia, and the total mass of CSM is at the level of about $10^{-4}\ \text{M}_{\odot}$ in agreement with the observations on the non-detection of H emission lines in the nebular spectrum (e.g., \citealt{2013MNRAS.435..329L,2016MNRAS.457.3254M,2018ApJ...863...24S,2020MNRAS.493.1044T}). 

\begin{table}
\begin{center}
\begin{tabular*}{0.83\linewidth}{cccccccc}
\hline
SNe  &  $R_1$ & $R_2$  & $R_3$    &  $\dot{M}_{\text{w}}(0)$  & $M_{\text{csm}}$/$10^{-4}\ \text{M}_{\odot}$  \\ 
\hline
12cg  & 6  & 8 & 20 &  3.0     &   0.52                     \\ 
14atg & 7   & 10    & 30 &  3.0     &    1.2                    \\ 
17cbv & 3   & 6   & 16 &  3.0   &   0.56                      \\ 
18oh  & 5   & 10   & 25 &  4.0     &   1.1                      \\ 
19np & 5 & 10  & 20 &  3.5    &  0.87                    \\ 
19yvq & 1.5 & 3  & 15 &  35.0     &  4.3                    \\ 
21aefx & 3 & 8  & 18 &  1.5    &  0.37                  \\ 
\hline
\end{tabular*} 
\caption{Here are the parameter values of Model\_ext for fitting the early flux excess of SNe~Ia except for SN~2020hvf. The unit of parameters $R_1$, $R_2$ and $R_3$ is $10^{14}$ cm, and that of $\dot{M}_{\text{w}}(0)$ is $10^{-6}\ \text{M}_{\odot}\cdot\text{yr}^{-1}$. $M_{\text{csm}}$ is the total mass of CSM integrated to the distance of $R_3$.} 
\label{para_model}
\end{center}
\end{table}

However, the great deviation of the predictions on $UVW1$-band luminosity suggests that the early-time excess of iPTF14atg may not be generated from the ejecta$-$CSM interaction but the ejecta$-$companion interaction since the ejecta-companion interaction can produce much higher temperature and hence more luminous UV-band radiation \citep{2010ApJ...708.1025K,2015Natur.521..328C,2016ApJ...820...92M}. As the discussion in \citet{2021ApJ...923L...8J}, the early excess of SN~2020hvf is highly possible to be generated from the CSM interaction process for its short duration of the optical flash. The values of parameter $R_{\text{out}}$ and $M_{\text{csm}}$ in fitting SN~2020hvf are slightly different from that in \citet{2021ApJ...923L...8J} due to the simplification of $L_{\text{opti}}$ for SN~2020hvf in this paper. The fitting $L_{\text{opti}}$ of SN~2018oh can only indicate that the ejecta$-$CSM interaction may be one of possible origination due to the lack of the early-time UV-band observations. For SNe~2012cg, 2017cbv, 2019np, and 2021aefx, the predicted $L_{UVW1}$ is consistent with the observed data considering the extinction from dusty CSM. A further diagnosis from radio observations is discussed in Section~\ref{section55}.

\section{The Radio Radiation from CSM Interaction} 
\label{section55}

An evident phenomenon of CSM interaction is the radio radiation emitted by the relativistic electrons. Although almost all the radio observations of spectroscopic normal SNe~Ia can only provide an upper limit, the radio radiation from ejecta$-$CSM interaction has important potential in distinguishing the various scenarios. The theory of the radio radiation from CSM interaction has been well established \citep{1982ApJ...259..302C,1998ApJ...499..810C,2014ApJ...787..143B,2014ApJ...792...38P,2020ApJ...890..159L}, and here we apply this theory to SNe~Ia with ejecta$-$CSM interaction soon after explosion. 

\subsection{The Synchrotron Radiation} 

A reasonable assumption is that the relativistic electrons produced by the ejecta$-$CSM interaction follow a power-law distribution, $\mathrm{d}N/\mathrm{d}E = N_0E^{-p}$, where $N$ and $N_0$ are the number density of the relativistic electrons and a scaling parameter, respectively. $E=\gamma m_e c^2$ is the energy of the electrons with $\gamma$ being the Lorentz factor. The corresponding synchrotron emission coefficient ($j_\nu$) is proportional to a declining power law of the frequency of the radiated photons, $j_\nu \propto \nu^{-\alpha}$, where the parameter $\alpha$ is equal to $(p-1)/2$. We adopt $\alpha = 1$ and $p = 3$ in this study.

\begin{figure*}
\centering
\includegraphics[width = 0.99 \linewidth]{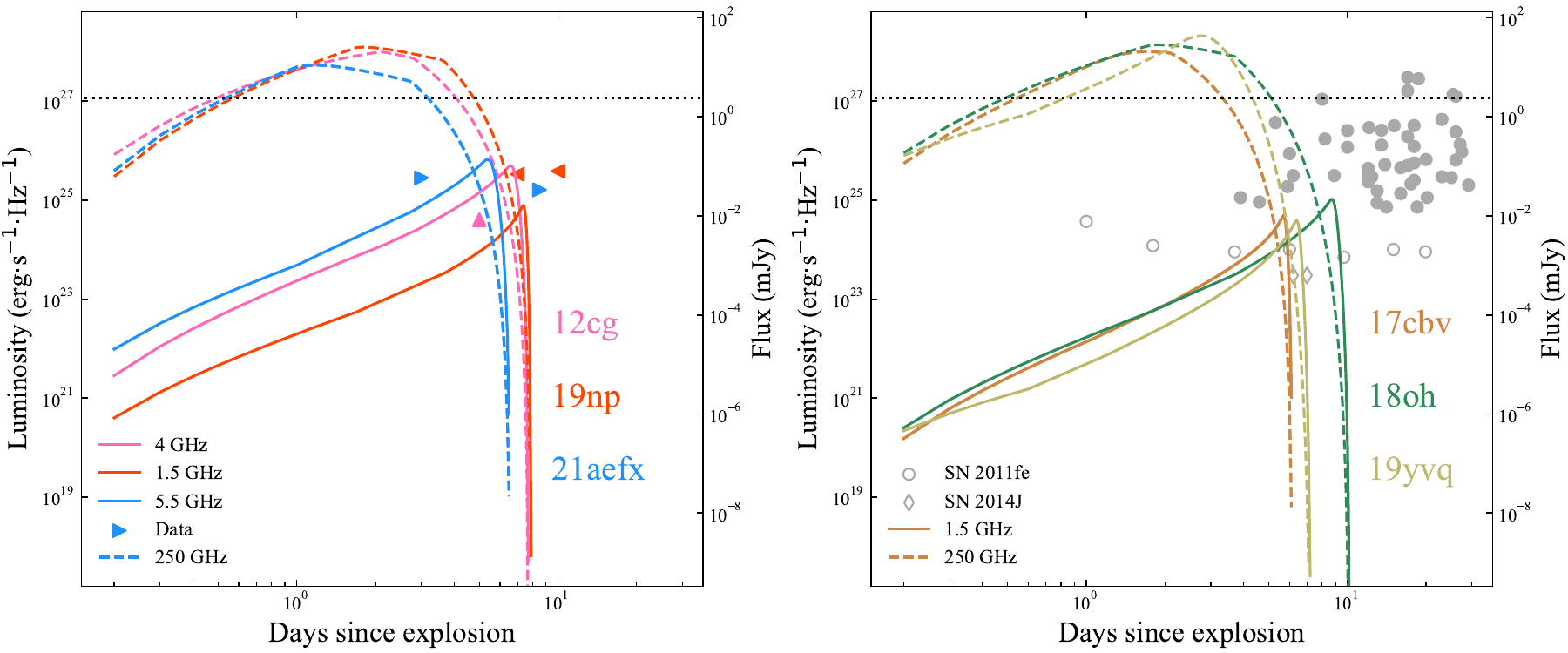}
\caption{The left panel: the symbols are the observed upper limits of radio fluxes of SN~2012cg (pink, 4.0 GHz), SN~2019np (purple, 1.5 GHz), and SN~2021aefx (blue, 5.5 GHz), respectively. The solid lines are the predicted radio luminosities derived from the ejecta-CSM interaction with the same parameter values shown in Table~\ref{para_model}. The right panel displays the predicted radio radiation in 1.5 GHz for SNe~2017cbv (solid orange line), 2018oh (solid green line), and 2019yvq (solid yellow line), comparing with the upper limits of radio fluxes of SNe~Ia (filled gray circles) from the tables in \citet{2016ApJ...821..119C}, including the highlighted SNe~2011fe (open circles) and 2014J (open diamond) and excluding the peculiar ones such as Iax, 02es-like, Ca-rich, super-Chandrasekhar, and Ia-CSM. The dashed lines in both panels are the corresponding radiation curves in 250 GHz. The corresponding radio flux scale assuming a distance of 20 Mpc, is displayed on the vertical axis on the right-hand side. For comparison, the horizontal dotted line is the sensitivity of ALMA with an integration time of 300 s. The synchrotron self-absorption effect is significant, especially at lower frequencies, and poses a challenge to radio observations at frequencies below $\sim10$ GHz.}
\label{fig_Figure_a5}
\end{figure*}

\subsection{The Synchrotron Self-Absorption} 

The effect of the synchrotron self-absorption (SSA) cannot be ignored because $N_0$ and the magnetic field ($B$) might be large enough to make the shocked area optically thick for the radio radiation. Assuming a uniform opacity distribution with the path length $\Delta s$, the optical depth $\tau_{\nu}$ is expressed as $\tau_{\nu} = \kappa_{\nu}\Delta s$, where $\kappa_{\nu}$ is the absorption coefficient and $\kappa_{\nu} = \kappa_0(p)N_0B^{(p+2)/2}\nu^{-(p+4)/2}$, where $\kappa_0(p)$ is a constant ($= 5.5\times10^{26}$ for $p = 3$). The intensity ($I_{\nu}$) is acquired by an integral as $I_{\nu} = \int_0^{\Delta s} j_{\nu}\exp(-\kappa_{\nu}s)\mathrm{d}s = \frac{j_{\nu}}{\kappa_{\nu}}(1-\exp(-\tau_{\nu}))$. Thus, the source function ($S_{\nu} = j_{\nu}/\kappa_{\nu}$) is proportional to $\nu^{5/2}$. 

For the simplicity of calculating $S_{\nu}$, we introduce a characteristic frequency $\nu_{\text{abs}}$, which has a corresponding optical depth $\tau_{\text{abs}} \sim 1$. This directly leads to $\tau_{\nu} = (\nu/\nu_{\text{abs}})^{-(p+4)/2}$. Besides, we can define a frequency $\nu_{\text{peak}}$ as $I_{\nu_{\text{peak}}}\equiv2kT_{\text{bright}}(\nu_{\text{peak}}/c)^2$, where $k$ is the Boltzmann constant and $T_{\text{bright}}$ is the brightness temperature. Thus, the intensity of any frequency can be formulated by $I_{\nu} = \frac{S_{\nu}}{S_{\nu_{\text{peak}}}}\frac{1-\exp(-\tau_{\nu})}{1-\exp(-\tau_{\nu_{\text{peak}}})}I_{\nu_{\text{peak}}}$. After proper arrangement, the formula is as follows, 
\begin{equation}
\label{eq_Inu}
I_{\nu} = \frac{2kT_{\text{bright}}}{c^2}\frac{\nu^{5/2}}{f(x)\nu_{\text{abs}}^{1/2}}[1-\exp(-\tau_{\nu})] 
\end{equation}
where $x = \nu_{\text{peak}}/\nu_{\text{abs}}$ and $f(x) = x^{1/2}[1 - \exp(-x^{-(p+4)/2})]$. Based on the Equation 12 in \cite{2014ApJ...787..143B}, $x \approx 1.137$ for $p = 3$. With $\tau_{\text{abs}}\sim 1$, we can get that $\nu_{\text{abs}} = (\Delta s\kappa_0(p)N_0B^{(p+2)/2})^{2/(p+4)}$.

\subsection{The Radio Luminosity from CSM Interaction} 

The kinetic evolution of the shocked shell can be constrained from Model\_ext with the assumption of $\Delta R_{\text{sh}} = 0.2R_{\text{sh}}$. Following the results in \citet{2014ApJ...792...38P}, we assume  that $\gamma_{\text{min}} \approx 1.64[V_{\text{sh}}/(70,000\ \text{km}\cdot{\rm s}^{-1})]^2 $ and $\gamma_{\text{min}} \ge 1$. We then have $N_0 = (p-2)\epsilon_{\text{rel}}u_{\text{th}}E_{\text{min}}^{p-2}$ by integrating the pow-law distribution of the relativistic electrons, where $u_{\text{th}} = (9/8)\rho_{\text{csm}}V_{\text{sh}}^2$ is the thermal energy density and $\epsilon_{\text{rel}}$ is the ratio of the energy density of the relativistic electrons and $u_{\text{th}}$. Besides, the magnetic field is determined by $B^2/(8\pi) = \epsilon_{B}u_{\text{th}}$, where $\epsilon_B$ is the ratio of the magnetic energy density and $u_{\text{th}}$. We set $\epsilon_{\text{rel}} = 0.1$ and $\epsilon_{B} = 0.01$ in our simulations \citep{2014ApJ...792...38P}.

Assuming that the shocked shell is homogeneous, the intensity along the line of sight is a function of the polar angle due to the path length. We define a parameter $h = \sin\theta$, where $\theta$ is the polar angle with respect to the direction of the line of sight. For $h = 0$, we denote $\nu_{\text{abs}} = \nu_{\text{abs,0}}$, $\tau_{\nu} = \tau_{\nu,0}$, and $\tau_{\nu_{\text{abs}}} = \tau_{\nu_{\text{abs,0}}} = 1$. For $0 \ge h \ge 1$, $\tau_{\nu}(h) = \xi_h\tau_{\nu,0}$, where $\xi_h = \Delta s(h)/(2\Delta R_{\text{sh}})$. Thus, $I_{\nu}(h)$ can be directly derived from Equation~\ref{eq_Inu} by replacing $\nu_{\text{abs}}$ and $\tau_{\nu}$ with $\nu_{\text{abs,0}}$ and $\tau_{\nu}(h)$, respectively. The luminosity $L_{\nu}$ is the integration over $h$ as $L_{\nu} = 8\pi^2R_{\text{sh}}^2\int_0^1I_{\nu}(h)h\mathrm{d}h$. We then define a factor $\vartheta = L_{\nu}/L_{\nu,0}$, where $L_{\nu,0} = 4\pi^2R_{\text{sh}}^2I_{\nu}(0)$. Thus we can get the observed luminosity as, 
\begin{equation}
\label{eq_Lv}
L_{\nu} = L_0\frac{\nu^{5/2}}{\nu_{\text{abs,0}}^{1/2}}[1-\exp(-\tau_{\nu,0})]
\end{equation}
where $L_0 = \frac{8\pi^2kT_{\text{bright}}}{c^2f(x)}R_{\text{sh}}^2\vartheta$. For optically thin or thick shell, Equation~\ref{eq_Lv} is reduced to $L_{\nu} = L_0\nu_{\text{abs,0}}^{(p+3)/2}\nu^{-(p-1)/2}$ or $L_{\nu} = L_0\nu^{5/2}/\nu_{\text{abs,0}}^{1/2}$, respectively. In our simulations, the optical depth $\tau_{\nu,0}$ evolves with time during the process of ejecta$-$CSM interaction. 

\subsection{The Predicted Radio Luminosity by the Model\_ext} 

Here, we compare the predicted radio radiation from the ejecta$-$CSM interaction process with the early-phase radio observations of SNe~Ia. 
For SNe~2012cg, 2019np, and 2021aefx, the predicted radio radiation is compared with their observed upper limits of radio fluxes. For SNe~2017cbv, 2018oh, and 2019yvq, the predicted radio radiation is compared with the observational data of normal SNe~Ia from \citet{2016ApJ...821..119C} due to the lack of the early-time radio observations of these three events. 
The predicted curves of radio luminosity for the low frequencies (e.g., 1.5 GHz, 4.0 GHz, and 5.5 GHz) and for the high frequency (250 GHz) are shown in Figure~\ref{fig_Figure_a5} with the same CSM parameter values as shown in Table\ref{para_model}. At the beginning of the ejecta$-$CSM interaction, the optical depth of radio bands is so large due to SSA that the radio luminosity at low frequencies is relatively low. As the shocked shell travels outwards, the CSM density rapidly decreases, resulting in a sharp decrease of the radio radiation for both high and low frequencies. The predicted radio luminosity is compared with the observations of SNe~Ia excluding the peculiar ones such as Iax, 02es-like, Ca-rich, super-Chandrasekhar, and Ia-CSM in Figure~\ref{fig_Figure_a5}. The predicted curves are below the upper limits of radio observations, except for SNe~2011fe and 2014J. This implies that even with the revisited SNe~Ia, which show obvious early light curve bumps, the existing observations are not sensitive enough to reveal the underlying CSM interaction. The progenitor mass loss rate soon before the explosion is even more tenuous for those SNe with detection limits lower than the predicted radio flux. For instance, the upper limit of the mass loss rate of SN~2011fe and SN~2014J is about $1\times10^{-10}\ \text{M}_{\odot}\cdot\text{yr}^{-1}$ from our calculation, which is a little bit smaller than the upper limit from \citet{2016ApJ...821..119C} due to the configuration setting of CSM interaction models.

Besides, the radio light curves shown in Figure~\ref{fig_Figure_a5} suggest that the radio observation at higher frequency (e.g., $\sim250\ \text{GHz}$) is several orders of magnitude stronger than at lower frequency (e.g., $\sim1.5\ \text{GHz}$). However, the biggest constraint is that the radio observations must be triggered within a few days after the explosion of SNe~Ia with early optical excess. Such high-frequency observations may be achievable by the Atacama Large Millimeter/submillimeter Array (ALMA) telescope. As shown in Figure~\ref{fig_Figure_a5}, the luminosity of $250\ \text{GHz}$ can exceed about $10^{27}\ \text{erg}\cdot{\rm s}^{-1}\cdot\text{Hz}^{-1}$ during $+1$ to $+5$ days respect to the explosion. The corresponding flux is about $10.0\ \text{mJy}$ at a  distance of about $20\ \text{Mpc}$, which happens to be within the sensitivity of the ALMA. It is critical to discover nearby SNe~Ia within one or two days after the explosion and triggering the multi-band photometric, spectral, and radio observations. The multi-messenger observations time-domain observational approach involving optical telescopes such as the Zwicky Transient Facility (ZTF, \citealt{2019PASP..131a8002B}), the Wide Field Survey Telescope (WFST, \citealt{2022Univ....9....7H,2023arXiv230607590W}), the Ultraviolet Transient Astronomy Satellite (ULTRASAT, \citealt{2022SPIE12181E..05B}), and ALMA radio observations will provide us the best chance to capture the UV, optical and radio signals from the ejecta$-$CSM interaction of SNe~Ia.

\section{Conclusions} 
\label{section66} 

In this paper, we revisited the possible ejecta$-$CSM interaction origin of the early excess emission in SNe~Ia. The CSM interaction described by Model\_sh is similar to that of the shock breakout process, in which the distance of CSM is about $10^{11}\sim10^{13}\ \text{cm}$. At such a short distance scale, the temperature of the shocked CSM rapidly decreases as it expands. Therefore, the corresponding thermal radiation duration is so short that Model\_sh can fit only the early flash of SN~2020hvf among the revisited eight SNe~Ia. When the radial distribution of CSM extends to about $10^{15}\ \text{cm}$, the CSM interaction continues for a few days. The Model\_ext describes a situation in which the mass-loss rate is a function of the time before the explosion. Under the appropriate parameter values, Model\_ext can fit the optical excess of the rest seven SNe~Ia. By considering the extinction and scattering from circumstellar dust, the Model\_ext can match the UV-band light curve except for iPTF14atg, which may rule out the possibility that the early excess emission in iPTF14atg arises from the ejecta$-$CSM interaction. In particular, the CSM interaction model relating to the case of Model\_ext also predicts radio radiations that can be detectable a few days past explosion at $\sim 250$ GHz, leading to a multi-band diagnosis of the circumstellar environment surrounding SNe~Ia. 

The success of Model\_ext in fitting the observed data of the revisited SNe~Ia suggests that the SNe~Ia with early excess require more observations to distinguish whether this excess originates from $^{56}$Ni mixing in the ejecta, Helium detonation on the surface of a WD, interaction with the companion, or ejecta$-$CSM interaction. It is necessary to compare the observational characteristics of these four scenarios in the first few days after the SN explosion. In particular, multi-messenger observations, including the X-ray, UV, optical, and radio bands, are all needed in distinguishing these scenarios. 

\section*{Acknowledgements}

This work is supported by the Major Science and Technology Project of Qinghai Province (2019-ZJ-A10) and the National Key Research and Development Programs of China (2022SKA0130100). Maokai Hu acknowledges support from the Jiangsu Funding Program for Excellent Postdoctoral Talent. Xiaofeng Wang is supported by the National Natural Science Foundation of China (NSFC grants NSFC grants 12288102, 12033003, and 11633002), the Scholar Program of Beijing Academy of Science and Technology (DZ: BS202002), and the Tencent Xplorer Prize. Lingzhi Wang is sponsored (in part) by the Chinese Academy of Sciences (CAS) through a grant to the CAS South America Center for Astronomy (CASSACA) in Santiago, Chile.

\section*{Data Availability}

The data underlying this article will be shared on reasonable request to the corresponding author.




\bibliographystyle{mnras}
\bibliography{example} 

\begin{thebibliography}{}
\makeatletter
\relax
\def\mn@urlcharsother{\let\do\@makeother \do\$\do\&\do\#\do\^\do\_\do\%\do\~}
\def\mn@doi{\begingroup\mn@urlcharsother \@ifnextchar [ {\mn@doi@}
  {\mn@doi@[]}}
\def\mn@doi@[#1]#2{\def\@tempa{#1}\ifx\@tempa\@empty \href
  {http://dx.doi.org/#2} {doi:#2}\else \href {http://dx.doi.org/#2} {#1}\fi
  \endgroup}
\def\mn@eprint#1#2{\mn@eprint@#1:#2::\@nil}
\def\mn@eprint@arXiv#1{\href {http://arxiv.org/abs/#1} {{\tt arXiv:#1}}}
\def\mn@eprint@dblp#1{\href {http://dblp.uni-trier.de/rec/bibtex/#1.xml}
  {dblp:#1}}
\def\mn@eprint@#1:#2:#3:#4\@nil{\def\@tempa {#1}\def\@tempb {#2}\def\@tempc
  {#3}\ifx \@tempc \@empty \let \@tempc \@tempb \let \@tempb \@tempa \fi \ifx
  \@tempb \@empty \def\@tempb {arXiv}\fi \@ifundefined
  {mn@eprint@\@tempb}{\@tempb:\@tempc}{\expandafter \expandafter \csname
  mn@eprint@\@tempb\endcsname \expandafter{\@tempc}}}

\bibitem[\protect\citeauthoryear{{Amanullah} \& {Goobar}}{{Amanullah} \&
  {Goobar}}{2011}]{2011ApJ...735...20A}
{Amanullah} R.,  {Goobar} A.,  2011, \mn@doi [\apj]
  {10.1088/0004-637X/735/1/20}, \href
  {https://ui.adsabs.harvard.edu/abs/2011ApJ...735...20A} {735, 20}

\bibitem[\protect\citeauthoryear{{Ashall} et~al.,}{{Ashall}
  et~al.}{2022}]{2022ApJ...932L...2A}
{Ashall} C.,  et~al., 2022, \mn@doi [\apjl] {10.3847/2041-8213/ac7235}, \href
  {https://ui.adsabs.harvard.edu/abs/2022ApJ...932L...2A} {932, L2}

\bibitem[\protect\citeauthoryear{{Bellm} et~al.,}{{Bellm}
  et~al.}{2019}]{2019PASP..131a8002B}
{Bellm} E.~C.,  et~al., 2019, \mn@doi [\pasp] {10.1088/1538-3873/aaecbe}, \href
  {https://ui.adsabs.harvard.edu/abs/2019PASP..131a8002B} {131, 018002}

\bibitem[\protect\citeauthoryear{{Ben-Ami} et~al.,}{{Ben-Ami}
  et~al.}{2022}]{2022SPIE12181E..05B}
{Ben-Ami} S.,  et~al., 2022, in {den Herder} J.-W.~A.,  {Nikzad} S.,
  {Nakazawa} K.,  eds,  Society of Photo-Optical Instrumentation Engineers
  (SPIE) Conference Series Vol. 12181, Space Telescopes and Instrumentation
  2022: Ultraviolet to Gamma Ray. p. 1218105 (\mn@eprint {arXiv} {2208.00159}),
  \mn@doi{10.1117/12.2629850}

\bibitem[\protect\citeauthoryear{{Bersten}, {Tanaka}, {Tominaga}, {Benvenuto}
  \& {Nomoto}}{{Bersten} et~al.}{2013}]{2013ApJ...767..143B}
{Bersten} M.~C.,  {Tanaka} M.,  {Tominaga} N.,  {Benvenuto} O.~G.,   {Nomoto}
  K.,  2013, \mn@doi [\apj] {10.1088/0004-637X/767/2/143}, \href
  {https://ui.adsabs.harvard.edu/abs/2013ApJ...767..143B} {767, 143}

\bibitem[\protect\citeauthoryear{{Bj{\"o}rnsson} \&
  {Lundqvist}}{{Bj{\"o}rnsson} \& {Lundqvist}}{2014}]{2014ApJ...787..143B}
{Bj{\"o}rnsson} C.~I.,  {Lundqvist} P.,  2014, \mn@doi [\apj]
  {10.1088/0004-637X/787/2/143}, \href
  {https://ui.adsabs.harvard.edu/abs/2014ApJ...787..143B} {787, 143}

\bibitem[\protect\citeauthoryear{{Bloom} et~al.,}{{Bloom}
  et~al.}{2012}]{2012ApJ...744L..17B}
{Bloom} J.~S.,  et~al., 2012, \mn@doi [\apjl] {10.1088/2041-8205/744/2/L17},
  \href {https://ui.adsabs.harvard.edu/abs/2012ApJ...744L..17B} {744, L17}

\bibitem[\protect\citeauthoryear{{Brown}, {Dawson}, {Harris}, {Olmstead},
  {Milne}  \& {Roming}}{{Brown} et~al.}{2012a}]{2012ApJ...749...18B}
{Brown} P.~J.,  {Dawson} K.~S.,  {Harris} D.~W.,  {Olmstead} M.,  {Milne} P.,
  {Roming} P. W.~A.,  2012a, \mn@doi [\apj] {10.1088/0004-637X/749/1/18}, \href
  {https://ui.adsabs.harvard.edu/abs/2012ApJ...749...18B} {749, 18}

\bibitem[\protect\citeauthoryear{{Brown} et~al.,}{{Brown}
  et~al.}{2012b}]{2012ApJ...753...22B}
{Brown} P.~J.,  et~al., 2012b, \mn@doi [\apj] {10.1088/0004-637X/753/1/22},
  \href {https://ui.adsabs.harvard.edu/abs/2012ApJ...753...22B} {753, 22}

\bibitem[\protect\citeauthoryear{{Bulla}, {Sim}, {Pakmor}, {Kromer},
  {Taubenberger}, {R{\"o}pke}, {Hillebrandt}  \& {Seitenzahl}}{{Bulla}
  et~al.}{2016}]{2016MNRAS.455.1060B}
{Bulla} M.,  {Sim} S.~A.,  {Pakmor} R.,  {Kromer} M.,  {Taubenberger} S.,
  {R{\"o}pke} F.~K.,  {Hillebrandt} W.,   {Seitenzahl} I.~R.,  2016, \mn@doi
  [MNRAS] {10.1093/mnras/stv2402}, \href
  {https://ui.adsabs.harvard.edu/abs/2016MNRAS.455.1060B} {455, 1060}

\bibitem[\protect\citeauthoryear{{Bulla} et~al.,}{{Bulla}
  et~al.}{2020}]{2020ApJ...902...48B}
{Bulla} M.,  et~al., 2020, \mn@doi [\apj] {10.3847/1538-4357/abb13c}, \href
  {https://ui.adsabs.harvard.edu/abs/2020ApJ...902...48B} {902, 48}

\bibitem[\protect\citeauthoryear{{Burke} et~al.,}{{Burke}
  et~al.}{2021}]{2021ApJ...919..142B}
{Burke} J.,  et~al., 2021, \mn@doi [\apj] {10.3847/1538-4357/ac126b}, \href
  {https://ui.adsabs.harvard.edu/abs/2021ApJ...919..142B} {919, 142}

\bibitem[\protect\citeauthoryear{{Burns} et~al.,}{{Burns}
  et~al.}{2020}]{2020ApJ...895..118B}
{Burns} C.~R.,  et~al., 2020, \mn@doi [\apj] {10.3847/1538-4357/ab8e3e}, \href
  {https://ui.adsabs.harvard.edu/abs/2020ApJ...895..118B} {895, 118}

\bibitem[\protect\citeauthoryear{{Cao} et~al.,}{{Cao}
  et~al.}{2015}]{2015Natur.521..328C}
{Cao} Y.,  et~al., 2015, \mn@doi [\nat] {10.1038/nature14440}, \href
  {https://ui.adsabs.harvard.edu/abs/2015Natur.521..328C} {521, 328}

\bibitem[\protect\citeauthoryear{{Chevalier}}{{Chevalier}}{1982a}]{1982ApJ...258..790C}
{Chevalier} R.~A.,  1982a, \mn@doi [\apj] {10.1086/160126}, \href
  {https://ui.adsabs.harvard.edu/abs/1982ApJ...258..790C} {258, 790}

\bibitem[\protect\citeauthoryear{{Chevalier}}{{Chevalier}}{1982b}]{1982ApJ...259..302C}
{Chevalier} R.~A.,  1982b, \mn@doi [\apj] {10.1086/160167}, \href
  {https://ui.adsabs.harvard.edu/abs/1982ApJ...259..302C} {259, 302}

\bibitem[\protect\citeauthoryear{{Chevalier}}{{Chevalier}}{1998}]{1998ApJ...499..810C}
{Chevalier} R.~A.,  1998, \mn@doi [\apj] {10.1086/305676}, \href
  {https://ui.adsabs.harvard.edu/abs/1998ApJ...499..810C} {499, 810}

\bibitem[\protect\citeauthoryear{{Chevalier} \& {Fransson}}{{Chevalier} \&
  {Fransson}}{1994}]{1994ApJ...420..268C}
{Chevalier} R.~A.,  {Fransson} C.,  1994, \mn@doi [\apj] {10.1086/173557},
  \href {https://ui.adsabs.harvard.edu/abs/1994ApJ...420..268C} {420, 268}

\bibitem[\protect\citeauthoryear{{Chomiuk} et~al.,}{{Chomiuk}
  et~al.}{2016}]{2016ApJ...821..119C}
{Chomiuk} L.,  et~al., 2016, \mn@doi [ApJ] {10.3847/0004-637X/821/2/119}, \href
  {https://ui.adsabs.harvard.edu/abs/2016ApJ...821..119C} {821, 119}

\bibitem[\protect\citeauthoryear{{Cikota} et~al.,}{{Cikota}
  et~al.}{2019}]{2019MNRAS.490..578C}
{Cikota} A.,  et~al., 2019, \mn@doi [MNRAS] {10.1093/mnras/stz2322}, \href
  {https://ui.adsabs.harvard.edu/abs/2019MNRAS.490..578C} {490, 578}

\bibitem[\protect\citeauthoryear{{Dekany} et~al.,}{{Dekany}
  et~al.}{2020}]{ZTF2020PASP..132c8001D}
{Dekany} R.,  et~al., 2020, \mn@doi [\pasp] {10.1088/1538-3873/ab4ca2}, \href
  {https://ui.adsabs.harvard.edu/abs/2020PASP..132c8001D} {132, 038001}

\bibitem[\protect\citeauthoryear{{Dimitriadis} et~al.,}{{Dimitriadis}
  et~al.}{2019}]{2019ApJ...870L...1D}
{Dimitriadis} G.,  et~al., 2019, \mn@doi [\apjl] {10.3847/2041-8213/aaedb0},
  \href {https://ui.adsabs.harvard.edu/abs/2019ApJ...870L...1D} {870, L1}

\bibitem[\protect\citeauthoryear{{Fausnaugh} et~al.,}{{Fausnaugh}
  et~al.}{2021}]{2021ApJ...908...51F}
{Fausnaugh} M.~M.,  et~al., 2021, \mn@doi [\apj] {10.3847/1538-4357/abcd42},
  \href {https://ui.adsabs.harvard.edu/abs/2021ApJ...908...51F} {908, 51}

\bibitem[\protect\citeauthoryear{{Filippenko}, {Li}, {Treffers}  \&
  {Modjaz}}{{Filippenko} et~al.}{2001}]{2001ASPC..246..121F}
{Filippenko} A.~V.,  {Li} W.~D.,  {Treffers} R.~R.,   {Modjaz} M.,  2001, in
  {Paczynski} B.,  {Chen} W.-P.,   {Lemme} C.,  eds,  Astronomical Society of
  the Pacific Conference Series Vol. 246, IAU Colloq. 183: Small Telescope
  Astronomy on Global Scales. p.~121

\bibitem[\protect\citeauthoryear{{Firth} et~al.,}{{Firth}
  et~al.}{2015}]{2015MNRAS.446.3895F}
{Firth} R.~E.,  et~al., 2015, \mn@doi [\mnras] {10.1093/mnras/stu2314}, \href
  {https://ui.adsabs.harvard.edu/abs/2015MNRAS.446.3895F} {446, 3895}

\bibitem[\protect\citeauthoryear{{F{\"o}rster} et~al.,}{{F{\"o}rster}
  et~al.}{2018}]{2018NatAs...2..808F}
{F{\"o}rster} F.,  et~al., 2018, \mn@doi [Nature Astronomy]
  {10.1038/s41550-018-0563-4}, \href
  {https://ui.adsabs.harvard.edu/abs/2018NatAs...2..808F} {2, 808}

\bibitem[\protect\citeauthoryear{{Ganeshalingam}, {Li}  \&
  {Filippenko}}{{Ganeshalingam} et~al.}{2011}]{2011MNRAS.416.2607G}
{Ganeshalingam} M.,  {Li} W.,   {Filippenko} A.~V.,  2011, \mn@doi [\mnras]
  {10.1111/j.1365-2966.2011.19213.x}, \href
  {https://ui.adsabs.harvard.edu/abs/2011MNRAS.416.2607G} {416, 2607}

\bibitem[\protect\citeauthoryear{{Graham} et~al.,}{{Graham}
  et~al.}{2019}]{2019PASP..131g8001G}
{Graham} M.~J.,  et~al., 2019, \mn@doi [PASP] {10.1088/1538-3873/ab006c}, \href
  {https://ui.adsabs.harvard.edu/abs/2019PASP..131g8001G} {131, 078001}

\bibitem[\protect\citeauthoryear{{Hillebrandt} \& {Niemeyer}}{{Hillebrandt} \&
  {Niemeyer}}{2000}]{2000ARA&A..38..191H}
{Hillebrandt} W.,  {Niemeyer} J.~C.,  2000, \mn@doi [\araa]
  {10.1146/annurev.astro.38.1.191}, \href
  {https://ui.adsabs.harvard.edu/abs/2000ARA&A..38..191H} {38, 191}

\bibitem[\protect\citeauthoryear{{Hosseinzadeh} et~al.,}{{Hosseinzadeh}
  et~al.}{2017}]{2017ApJ...845L..11H}
{Hosseinzadeh} G.,  et~al., 2017, \mn@doi [\apjl] {10.3847/2041-8213/aa8402},
  \href {https://ui.adsabs.harvard.edu/abs/2017ApJ...845L..11H} {845, L11}

\bibitem[\protect\citeauthoryear{{Hosseinzadeh} et~al.,}{{Hosseinzadeh}
  et~al.}{2022}]{2022ApJ...933L..45H}
{Hosseinzadeh} G.,  et~al., 2022, \mn@doi [\apjl] {10.3847/2041-8213/ac7cef},
  \href {https://ui.adsabs.harvard.edu/abs/2022ApJ...933L..45H} {933, L45}

\bibitem[\protect\citeauthoryear{{Howell}}{{Howell}}{2011}]{2011NatCo...2..350H}
{Howell} D.~A.,  2011, \mn@doi [Nature Communications] {10.1038/ncomms1344},
  \href {https://ui.adsabs.harvard.edu/abs/2011NatCo...2..350H} {2, 350}

\bibitem[\protect\citeauthoryear{{Hu}, {Hu}, {Jiang}, {Xiao}, {Fan}, {Wei}  \&
  {Wu}}{{Hu} et~al.}{2022a}]{2022Univ....9....7H}
{Hu} M.,  {Hu} L.,  {Jiang} J.-a.,  {Xiao} L.,  {Fan} L.,  {Wei} J.,   {Wu} X.,
   2022a, \mn@doi [Universe] {10.3390/universe9010007}, \href
  {https://ui.adsabs.harvard.edu/abs/2022Univ....9....7H} {9, 7}

\bibitem[\protect\citeauthoryear{{Hu}, {Wang}  \& {Wang}}{{Hu}
  et~al.}{2022b}]{2022ApJ...931..110H}
{Hu} M.,  {Wang} L.,   {Wang} X.,  2022b, \mn@doi [\apj]
  {10.3847/1538-4357/ac6be5}, \href
  {https://ui.adsabs.harvard.edu/abs/2022ApJ...931..110H} {931, 110}

\bibitem[\protect\citeauthoryear{{Iben} \& {Tutukov}}{{Iben} \&
  {Tutukov}}{1984}]{1984ApJS...54..335I}
{Iben} I. J.,  {Tutukov} A.~V.,  1984, \mn@doi [\apjs] {10.1086/190932}, \href
  {https://ui.adsabs.harvard.edu/abs/1984ApJS...54..335I} {54, 335}

\bibitem[\protect\citeauthoryear{{Jiang} et~al.,}{{Jiang}
  et~al.}{2017}]{2017Natur.550...80J}
{Jiang} J.-A.,  et~al., 2017, \mn@doi [\nat] {10.1038/nature23908}, \href
  {https://ui.adsabs.harvard.edu/abs/2017Natur.550...80J} {550, 80}

\bibitem[\protect\citeauthoryear{{Jiang}, {Doi}, {Maeda}  \&
  {Shigeyama}}{{Jiang} et~al.}{2018}]{2018ApJ...865..149J}
{Jiang} J.-a.,  {Doi} M.,  {Maeda} K.,   {Shigeyama} T.,  2018, \mn@doi [\apj]
  {10.3847/1538-4357/aadb9a}, \href
  {https://ui.adsabs.harvard.edu/abs/2018ApJ...865..149J} {865, 149}

\bibitem[\protect\citeauthoryear{{Jiang} et~al.,}{{Jiang}
  et~al.}{2020}]{2020ApJ...892...25J}
{Jiang} J.-a.,  et~al., 2020, \mn@doi [\apj] {10.3847/1538-4357/ab76cb}, \href
  {https://ui.adsabs.harvard.edu/abs/2020ApJ...892...25J} {892, 25}

\bibitem[\protect\citeauthoryear{{Jiang} et~al.,}{{Jiang}
  et~al.}{2021}]{2021ApJ...923L...8J}
{Jiang} J.-a.,  et~al., 2021, \mn@doi [\apjl] {10.3847/2041-8213/ac375f}, \href
  {https://ui.adsabs.harvard.edu/abs/2021ApJ...923L...8J} {923, L8}

\bibitem[\protect\citeauthoryear{{Jin}, {Yoon}  \& {Blinnikov}}{{Jin}
  et~al.}{2021}]{2021ApJ...910...68J}
{Jin} H.,  {Yoon} S.-C.,   {Blinnikov} S.,  2021, \mn@doi [\apj]
  {10.3847/1538-4357/abe0b1}, \href
  {https://ui.adsabs.harvard.edu/abs/2021ApJ...910...68J} {910, 68}

\bibitem[\protect\citeauthoryear{{Kasen}}{{Kasen}}{2010}]{2010ApJ...708.1025K}
{Kasen} D.,  2010, \mn@doi [\apj] {10.1088/0004-637X/708/2/1025}, \href
  {https://ui.adsabs.harvard.edu/abs/2010ApJ...708.1025K} {708, 1025}

\bibitem[\protect\citeauthoryear{{Kilpatrick} et~al.,}{{Kilpatrick}
  et~al.}{2018}]{2018MNRAS.481.4123K}
{Kilpatrick} C.~D.,  et~al., 2018, \mn@doi [MNRAS] {10.1093/mnras/sty2503},
  \href {https://ui.adsabs.harvard.edu/abs/2018MNRAS.481.4123K} {481, 4123}

\bibitem[\protect\citeauthoryear{{Kochanek} et~al.,}{{Kochanek}
  et~al.}{2017}]{2017PASP..129j4502K}
{Kochanek} C.~S.,  et~al., 2017, \mn@doi [PASP] {10.1088/1538-3873/aa80d9},
  \href {https://ui.adsabs.harvard.edu/abs/2017PASP..129j4502K} {129, 104502}

\bibitem[\protect\citeauthoryear{{Law} et~al.,}{{Law}
  et~al.}{2009}]{2009PASP..121.1395L}
{Law} N.~M.,  et~al., 2009, \mn@doi [PASP] {10.1086/648598}, \href
  {https://ui.adsabs.harvard.edu/abs/2009PASP..121.1395L} {121, 1395}

\bibitem[\protect\citeauthoryear{{Levanon} \& {Soker}}{{Levanon} \&
  {Soker}}{2019}]{2019ApJ...872L...7L}
{Levanon} N.,  {Soker} N.,  2019, \mn@doi [\apjl] {10.3847/2041-8213/ab0285},
  \href {https://ui.adsabs.harvard.edu/abs/2019ApJ...872L...7L} {872, L7}

\bibitem[\protect\citeauthoryear{{Li} et~al.,}{{Li}
  et~al.}{2011}]{2011Natur.480..348L}
{Li} W.,  et~al., 2011, \mn@doi [\nat] {10.1038/nature10646}, \href
  {https://ui.adsabs.harvard.edu/abs/2011Natur.480..348L} {480, 348}

\bibitem[\protect\citeauthoryear{{Li} et~al.,}{{Li}
  et~al.}{2019}]{2019ApJ...870...12L}
{Li} W.,  et~al., 2019, \mn@doi [\apj] {10.3847/1538-4357/aaec74}, \href
  {https://ui.adsabs.harvard.edu/abs/2019ApJ...870...12L} {870, 12}

\bibitem[\protect\citeauthoryear{{Li} et~al.,}{{Li}
  et~al.}{2021}]{2021ApJ...906...99L}
{Li} W.,  et~al., 2021, \mn@doi [\apj] {10.3847/1538-4357/abc9b5}, \href
  {https://ui.adsabs.harvard.edu/abs/2021ApJ...906...99L} {906, 99}

\bibitem[\protect\citeauthoryear{{Lundqvist} et~al.,}{{Lundqvist}
  et~al.}{2013}]{2013MNRAS.435..329L}
{Lundqvist} P.,  et~al., 2013, \mn@doi [MNRAS] {10.1093/mnras/stt1303}, \href
  {https://ui.adsabs.harvard.edu/abs/2013MNRAS.435..329L} {435, 329}

\bibitem[\protect\citeauthoryear{{Lundqvist} et~al.,}{{Lundqvist}
  et~al.}{2020}]{2020ApJ...890..159L}
{Lundqvist} P.,  et~al., 2020, \mn@doi [ApJ] {10.3847/1538-4357/ab6dc6}, \href
  {https://ui.adsabs.harvard.edu/abs/2020ApJ...890..159L} {890, 159}

\bibitem[\protect\citeauthoryear{{Maeda}, {Kutsuna}  \& {Shigeyama}}{{Maeda}
  et~al.}{2014}]{2014ApJ...794...37M}
{Maeda} K.,  {Kutsuna} M.,   {Shigeyama} T.,  2014, \mn@doi [\apj]
  {10.1088/0004-637X/794/1/37}, \href
  {https://ui.adsabs.harvard.edu/abs/2014ApJ...794...37M} {794, 37}

\bibitem[\protect\citeauthoryear{{Maeda}, {Jiang}, {Shigeyama}  \&
  {Doi}}{{Maeda} et~al.}{2018}]{2018ApJ...861...78M}
{Maeda} K.,  {Jiang} J.-a.,  {Shigeyama} T.,   {Doi} M.,  2018, \mn@doi [\apj]
  {10.3847/1538-4357/aac8d8}, \href
  {https://ui.adsabs.harvard.edu/abs/2018ApJ...861...78M} {861, 78}

\bibitem[\protect\citeauthoryear{{Magee} \& {Maguire}}{{Magee} \&
  {Maguire}}{2020}]{2020A&A...642A.189M}
{Magee} M.~R.,  {Maguire} K.,  2020, \mn@doi [\aap]
  {10.1051/0004-6361/202037870}, \href
  {https://ui.adsabs.harvard.edu/abs/2020A&A...642A.189M} {642, A189}

\bibitem[\protect\citeauthoryear{{Magee}, {Sim}, {Kotak}  \&
  {Kerzendorf}}{{Magee} et~al.}{2018}]{2018A&A...614A.115M}
{Magee} M.~R.,  {Sim} S.~A.,  {Kotak} R.,   {Kerzendorf} W.~E.,  2018, \mn@doi
  [\aap] {10.1051/0004-6361/201832675}, \href
  {https://ui.adsabs.harvard.edu/abs/2018A&A...614A.115M} {614, A115}

\bibitem[\protect\citeauthoryear{{Magee}, {Maguire}, {Kotak}, {Sim},
  {Gillanders}, {Prentice}  \& {Skillen}}{{Magee}
  et~al.}{2020}]{2020A&A...634A..37M}
{Magee} M.~R.,  {Maguire} K.,  {Kotak} R.,  {Sim} S.~A.,  {Gillanders} J.~H.,
  {Prentice} S.~J.,   {Skillen} K.,  2020, \mn@doi [\aap]
  {10.1051/0004-6361/201936684}, \href
  {https://ui.adsabs.harvard.edu/abs/2020A&A...634A..37M} {634, A37}

\bibitem[\protect\citeauthoryear{{Magee}, {Maguire}, {Kotak}  \& {Sim}}{{Magee}
  et~al.}{2021}]{2021MNRAS.502.3533M}
{Magee} M.~R.,  {Maguire} K.,  {Kotak} R.,   {Sim} S.~A.,  2021, \mn@doi
  [\mnras] {10.1093/mnras/stab201}, \href
  {https://ui.adsabs.harvard.edu/abs/2021MNRAS.502.3533M} {502, 3533}

\bibitem[\protect\citeauthoryear{{Maguire}, {Taubenberger}, {Sullivan}  \&
  {Mazzali}}{{Maguire} et~al.}{2016}]{2016MNRAS.457.3254M}
{Maguire} K.,  {Taubenberger} S.,  {Sullivan} M.,   {Mazzali} P.~A.,  2016,
  \mn@doi [MNRAS] {10.1093/mnras/stv2991}, \href
  {https://ui.adsabs.harvard.edu/abs/2016MNRAS.457.3254M} {457, 3254}

\bibitem[\protect\citeauthoryear{{Maoz}, {Mannucci}  \& {Nelemans}}{{Maoz}
  et~al.}{2014}]{2014ARA&A..52..107M}
{Maoz} D.,  {Mannucci} F.,   {Nelemans} G.,  2014, \mn@doi [ARA\&A]
  {10.1146/annurev-astro-082812-141031}, \href
  {https://ui.adsabs.harvard.edu/abs/2014ARA&A..52..107M} {52, 107}

\bibitem[\protect\citeauthoryear{{Marion} et~al.,}{{Marion}
  et~al.}{2016}]{2016ApJ...820...92M}
{Marion} G.~H.,  et~al., 2016, \mn@doi [\apj] {10.3847/0004-637X/820/2/92},
  \href {https://ui.adsabs.harvard.edu/abs/2016ApJ...820...92M} {820, 92}

\bibitem[\protect\citeauthoryear{{Mattila}, {Lundqvist}, {Sollerman}, {Kozma},
  {Baron}, {Fransson}, {Leibundgut}  \& {Nomoto}}{{Mattila}
  et~al.}{2005}]{2005A&A...443..649M}
{Mattila} S.,  {Lundqvist} P.,  {Sollerman} J.,  {Kozma} C.,  {Baron} E.,
  {Fransson} C.,  {Leibundgut} B.,   {Nomoto} K.,  2005, \mn@doi [A\&A]
  {10.1051/0004-6361:20052731}, \href
  {https://ui.adsabs.harvard.edu/abs/2005A&A...443..649M} {443, 649}

\bibitem[\protect\citeauthoryear{{Matzner} \& {McKee}}{{Matzner} \&
  {McKee}}{1999}]{1999ApJ...510..379M}
{Matzner} C.~D.,  {McKee} C.~F.,  1999, \mn@doi [\apj] {10.1086/306571}, \href
  {https://ui.adsabs.harvard.edu/abs/1999ApJ...510..379M} {510, 379}

\bibitem[\protect\citeauthoryear{{Miller} et~al.,}{{Miller}
  et~al.}{2020a}]{2020ApJ...898...56M}
{Miller} A.~A.,  et~al., 2020a, \mn@doi [\apj] {10.3847/1538-4357/ab9e05},
  \href {https://ui.adsabs.harvard.edu/abs/2020ApJ...898...56M} {898, 56}

\bibitem[\protect\citeauthoryear{{Miller} et~al.,}{{Miller}
  et~al.}{2020b}]{2020ApJ...902...47M}
{Miller} A.~A.,  et~al., 2020b, \mn@doi [\apj] {10.3847/1538-4357/abb13b},
  \href {https://ui.adsabs.harvard.edu/abs/2020ApJ...902...47M} {902, 47}

\bibitem[\protect\citeauthoryear{{Moriya}, {Maeda}, {Taddia}, {Sollerman},
  {Blinnikov}  \& {Sorokina}}{{Moriya} et~al.}{2013}]{2013MNRAS.435.1520M}
{Moriya} T.~J.,  {Maeda} K.,  {Taddia} F.,  {Sollerman} J.,  {Blinnikov} S.~I.,
    {Sorokina} E.~I.,  2013, \mn@doi [\mnras] {10.1093/mnras/stt1392}, \href
  {https://ui.adsabs.harvard.edu/abs/2013MNRAS.435.1520M} {435, 1520}

\bibitem[\protect\citeauthoryear{{Moriya}, {Mazzali}, {Ashall}  \&
  {Pian}}{{Moriya} et~al.}{2023}]{2023MNRAS.522.6035M}
{Moriya} T.~J.,  {Mazzali} P.~A.,  {Ashall} C.,   {Pian} E.,  2023, \mn@doi
  [\mnras] {10.1093/mnras/stad1386}, \href
  {https://ui.adsabs.harvard.edu/abs/2023MNRAS.522.6035M} {522, 6035}

\bibitem[\protect\citeauthoryear{{Nelemans}, {Voss}, {Roelofs}  \&
  {Bassa}}{{Nelemans} et~al.}{2008}]{2008MNRAS.388..487N}
{Nelemans} G.,  {Voss} R.,  {Roelofs} G.,   {Bassa} C.,  2008, \mn@doi [MNRAS]
  {10.1111/j.1365-2966.2008.13416.x}, \href
  {https://ui.adsabs.harvard.edu/abs/2008MNRAS.388..487N} {388, 487}

\bibitem[\protect\citeauthoryear{{Nomoto}}{{Nomoto}}{1982}]{1982ApJ...253..798N}
{Nomoto} K.,  1982, \mn@doi [\apj] {10.1086/159682}, \href
  {https://ui.adsabs.harvard.edu/abs/1982ApJ...253..798N} {253, 798}

\bibitem[\protect\citeauthoryear{{Nugent} et~al.,}{{Nugent}
  et~al.}{2011}]{2011Natur.480..344N}
{Nugent} P.~E.,  et~al., 2011, \mn@doi [\nat] {10.1038/nature10644}, \href
  {https://ui.adsabs.harvard.edu/abs/2011Natur.480..344N} {480, 344}

\bibitem[\protect\citeauthoryear{{Panagia}, {Van Dyk}, {Weiler}, {Sramek},
  {Stockdale}  \& {Murata}}{{Panagia} et~al.}{2006}]{2006ApJ...646..369P}
{Panagia} N.,  {Van Dyk} S.~D.,  {Weiler} K.~W.,  {Sramek} R.~A.,  {Stockdale}
  C.~J.,   {Murata} K.~P.,  2006, \mn@doi [ApJ] {10.1086/504710}, \href
  {https://ui.adsabs.harvard.edu/abs/2006ApJ...646..369P} {646, 369}

\bibitem[\protect\citeauthoryear{{Pereira} et~al.,}{{Pereira}
  et~al.}{2013}]{2013A&A...554A..27P}
{Pereira} R.,  et~al., 2013, \mn@doi [\aap] {10.1051/0004-6361/201221008},
  \href {https://ui.adsabs.harvard.edu/abs/2013A&A...554A..27P} {554, A27}

\bibitem[\protect\citeauthoryear{{P{\'e}rez-Torres} et~al.,}{{P{\'e}rez-Torres}
  et~al.}{2014}]{2014ApJ...792...38P}
{P{\'e}rez-Torres} M.~A.,  et~al., 2014, \mn@doi [\apj]
  {10.1088/0004-637X/792/1/38}, \href
  {https://ui.adsabs.harvard.edu/abs/2014ApJ...792...38P} {792, 38}

\bibitem[\protect\citeauthoryear{{Perlmutter} et~al.,}{{Perlmutter}
  et~al.}{1999}]{1999ApJ...517..565P}
{Perlmutter} S.,  et~al., 1999, \mn@doi [\apj] {10.1086/307221}, \href
  {https://ui.adsabs.harvard.edu/abs/1999ApJ...517..565P} {517, 565}

\bibitem[\protect\citeauthoryear{{Piro}}{{Piro}}{2015}]{2015ApJ...808L..51P}
{Piro} A.~L.,  2015, \mn@doi [\apjl] {10.1088/2041-8205/808/2/L51}, \href
  {https://ui.adsabs.harvard.edu/abs/2015ApJ...808L..51P} {808, L51}

\bibitem[\protect\citeauthoryear{{Porter} et~al.,}{{Porter}
  et~al.}{2016}]{2016ApJ...828...24P}
{Porter} A.~L.,  et~al., 2016, \mn@doi [ApJ] {10.3847/0004-637X/828/1/24},
  \href {https://ui.adsabs.harvard.edu/abs/2016ApJ...828...24P} {828, 24}

\bibitem[\protect\citeauthoryear{{Riess} et~al.,}{{Riess}
  et~al.}{1998}]{1998AJ....116.1009R}
{Riess} A.~G.,  et~al., 1998, \mn@doi [\aj] {10.1086/300499}, \href
  {https://ui.adsabs.harvard.edu/abs/1998AJ....116.1009R} {116, 1009}

\bibitem[\protect\citeauthoryear{{Riess} et~al.,}{{Riess}
  et~al.}{1999}]{1999AJ....118.2675R}
{Riess} A.~G.,  et~al., 1999, \mn@doi [\aj] {10.1086/301143}, \href
  {https://ui.adsabs.harvard.edu/abs/1999AJ....118.2675R} {118, 2675}

\bibitem[\protect\citeauthoryear{{Riess} et~al.,}{{Riess}
  et~al.}{2007}]{2007ApJ...659...98R}
{Riess} A.~G.,  et~al., 2007, \mn@doi [\apj] {10.1086/510378}, \href
  {https://ui.adsabs.harvard.edu/abs/2007ApJ...659...98R} {659, 98}

\bibitem[\protect\citeauthoryear{{Russell} \& {Immler}}{{Russell} \&
  {Immler}}{2012}]{2012ApJ...748L..29R}
{Russell} B.~R.,  {Immler} S.,  2012, \mn@doi [ApJL]
  {10.1088/2041-8205/748/2/L29}, \href
  {https://ui.adsabs.harvard.edu/abs/2012ApJ...748L..29R} {748, L29}

\bibitem[\protect\citeauthoryear{{Sai} et~al.,}{{Sai}
  et~al.}{2022}]{2022MNRAS.514.3541S}
{Sai} H.,  et~al., 2022, \mn@doi [\mnras] {10.1093/mnras/stac1525}, \href
  {https://ui.adsabs.harvard.edu/abs/2022MNRAS.514.3541S} {514, 3541}

\bibitem[\protect\citeauthoryear{{Sand} et~al.,}{{Sand}
  et~al.}{2018}]{2018ApJ...863...24S}
{Sand} D.~J.,  et~al., 2018, \mn@doi [\apj] {10.3847/1538-4357/aacde8}, \href
  {https://ui.adsabs.harvard.edu/abs/2018ApJ...863...24S} {863, 24}

\bibitem[\protect\citeauthoryear{{Shappee}, {Stanek}, {Pogge}  \&
  {Garnavich}}{{Shappee} et~al.}{2013}]{2013ApJ...762L...5S}
{Shappee} B.~J.,  {Stanek} K.~Z.,  {Pogge} R.~W.,   {Garnavich} P.~M.,  2013,
  \mn@doi [ApJL] {10.1088/2041-8205/762/1/L5}, \href
  {https://ui.adsabs.harvard.edu/abs/2013ApJ...762L...5S} {762, L5}

\bibitem[\protect\citeauthoryear{{Svirski}, {Nakar}  \& {Sari}}{{Svirski}
  et~al.}{2012}]{2012ApJ...759..108S}
{Svirski} G.,  {Nakar} E.,   {Sari} R.,  2012, \mn@doi [\apj]
  {10.1088/0004-637X/759/2/108}, \href
  {https://ui.adsabs.harvard.edu/abs/2012ApJ...759..108S} {759, 108}

\bibitem[\protect\citeauthoryear{{Takei} \& {Shigeyama}}{{Takei} \&
  {Shigeyama}}{2020}]{2020PASJ...72...67T}
{Takei} Y.,  {Shigeyama} T.,  2020, \mn@doi [\pasj] {10.1093/pasj/psaa050},
  \href {https://ui.adsabs.harvard.edu/abs/2020PASJ...72...67T} {72, 67}

\bibitem[\protect\citeauthoryear{{Tonry} et~al.,}{{Tonry}
  et~al.}{2018}]{ATLAS2018PASP..130f4505T}
{Tonry} J.~L.,  et~al., 2018, \mn@doi [\pasp] {10.1088/1538-3873/aabadf}, \href
  {https://ui.adsabs.harvard.edu/abs/2018PASP..130f4505T} {130, 064505}

\bibitem[\protect\citeauthoryear{{Tucker} et~al.,}{{Tucker}
  et~al.}{2020}]{2020MNRAS.493.1044T}
{Tucker} M.~A.,  et~al., 2020, \mn@doi [MNRAS] {10.1093/mnras/stz3390}, \href
  {https://ui.adsabs.harvard.edu/abs/2020MNRAS.493.1044T} {493, 1044}

\bibitem[\protect\citeauthoryear{{WFST Collaboration} et~al.,}{{WFST
  Collaboration} et~al.}{2023}]{2023arXiv230607590W}
{WFST Collaboration} et~al., 2023, \mn@doi [arXiv e-prints]
  {10.48550/arXiv.2306.07590}, \href
  {https://ui.adsabs.harvard.edu/abs/2023arXiv230607590W} {p. arXiv:2306.07590}

\bibitem[\protect\citeauthoryear{Wang \& Wheeler}{Wang \&
  Wheeler}{2008}]{doi:10.1146/annurev.astro.46.060407.145139}
Wang L.,  Wheeler J.~C.,  2008, \mn@doi [Annual Review of Astronomy and
  Astrophysics] {10.1146/annurev.astro.46.060407.145139}, 46, 433

\bibitem[\protect\citeauthoryear{{Wang}, {Wheeler}  \& {H{\"o}flich}}{{Wang}
  et~al.}{1997}]{1997ApJ...476L..27W}
{Wang} L.,  {Wheeler} J.~C.,   {H{\"o}flich} P.,  1997, \mn@doi [ApJL]
  {10.1086/310495}, \href
  {https://ui.adsabs.harvard.edu/abs/1997ApJ...476L..27W} {476, L27}

\bibitem[\protect\citeauthoryear{{Wang}, {Wang}, {Filippenko}, {Zhang}  \&
  {Zhao}}{{Wang} et~al.}{2013}]{2013Sci...340..170W}
{Wang} X.,  {Wang} L.,  {Filippenko} A.~V.,  {Zhang} T.,   {Zhao} X.,  2013,
  \mn@doi [Science] {10.1126/science.1231502}, \href
  {https://ui.adsabs.harvard.edu/abs/2013Sci...340..170W} {340, 170}

\bibitem[\protect\citeauthoryear{{Wang}, {Chen}, {Wang}, {Hu}, {Xi}, {Yang},
  {Zhao}  \& {Li}}{{Wang} et~al.}{2019}]{2019ApJ...882..120W}
{Wang} X.,  {Chen} J.,  {Wang} L.,  {Hu} M.,  {Xi} G.,  {Yang} Y.,  {Zhao} X.,
   {Li} W.,  2019, \mn@doi [\apj] {10.3847/1538-4357/ab26b5}, \href
  {https://ui.adsabs.harvard.edu/abs/2019ApJ...882..120W} {882, 120}

\bibitem[\protect\citeauthoryear{{Wang} et~al.,}{{Wang}
  et~al.}{2020}]{2020ApJ...904...14W}
{Wang} L.,  et~al., 2020, \mn@doi [\apj] {10.3847/1538-4357/abba82}, \href
  {https://ui.adsabs.harvard.edu/abs/2020ApJ...904...14W} {904, 14}

\bibitem[\protect\citeauthoryear{{Webbink}}{{Webbink}}{1984}]{1984ApJ...277..355W}
{Webbink} R.~F.,  1984, \mn@doi [\apj] {10.1086/161701}, \href
  {https://ui.adsabs.harvard.edu/abs/1984ApJ...277..355W} {277, 355}

\bibitem[\protect\citeauthoryear{{Wee}, {Chakraborty}, {Wang}  \&
  {Penprase}}{{Wee} et~al.}{2018}]{2018ApJ...863...90W}
{Wee} J.,  {Chakraborty} N.,  {Wang} J.,   {Penprase} B.~E.,  2018, \mn@doi
  [\apj] {10.3847/1538-4357/aacd4e}, \href
  {https://ui.adsabs.harvard.edu/abs/2018ApJ...863...90W} {863, 90}

\bibitem[\protect\citeauthoryear{{Whelan} \& {Iben}}{{Whelan} \&
  {Iben}}{1973}]{1973ApJ...186.1007W}
{Whelan} J.,  {Iben} Icko J.,  1973, \mn@doi [\apj] {10.1086/152565}, \href
  {https://ui.adsabs.harvard.edu/abs/1973ApJ...186.1007W} {186, 1007}

\bibitem[\protect\citeauthoryear{{Wood-Vasey}, {Wang}  \&
  {Aldering}}{{Wood-Vasey} et~al.}{2004}]{Wood-Vasey:2004ApJ...616..339W}
{Wood-Vasey} W.~M.,  {Wang} L.,   {Aldering} G.,  2004, \mn@doi [ApJ]
  {10.1086/424826}, \href
  {https://ui.adsabs.harvard.edu/abs/2004ApJ...616..339W} {616, 339}

\bibitem[\protect\citeauthoryear{{Yang} et~al.,}{{Yang}
  et~al.}{2020}]{2020ApJ...902...46Y}
{Yang} Y.,  et~al., 2020, \mn@doi [ApJ] {10.3847/1538-4357/aba759}, \href
  {https://ui.adsabs.harvard.edu/abs/2020ApJ...902...46Y} {902, 46}

\bibitem[\protect\citeauthoryear{{Zhang} et~al.,}{{Zhang}
  et~al.}{2016}]{2016ApJ...820...67Z}
{Zhang} K.,  et~al., 2016, \mn@doi [\apj] {10.3847/0004-637X/820/1/67}, \href
  {https://ui.adsabs.harvard.edu/abs/2016ApJ...820...67Z} {820, 67}

\makeatother
\end{thebibliography}








\bsp	
\label{lastpage}
\end{document}